%% file: paper.tex
\documentclass[twocolumn,showpacs,aps,floatfix,prd]{revtex4}

\usepackage{graphicx}
\usepackage{dcolumn}
\usepackage{amsmath}
\usepackage{epsfig}
\usepackage{hhline}
\usepackage{soul}
\usepackage{tabularx,pifont,multirow}
\usepackage{enumitem}
\usepackage{colordvi}

\RequirePackage{xspace}

\newcommand{\BABARPubYear}    {04}
\newcommand{\BABARPubNumber}  {001}
\newcommand{\SLACPubNumber} {10364}

\input pubboard/babarsym
\input symbols

\input symbolscpt

\newcommand{\pvec}{{\bf p}} 

\newcommand{\ket}[1]{\ensuremath{|{#1}\rangle}}
  
\def\beq{\begin{equation}}
\def\eeq{\end{equation}}
\def\bea{\begin{eqnarray}}
\def\eea{\end{eqnarray}}
\def\bq{\begin{quote}}
\def\eq{\end{quote}}
\def\ben{\begin{enumerate}}
\def\een{\end{enumerate}}
\def\nn{\nonumber}

\newcommand{\pham}{\ensuremath{\phantom{-}}}
\newcommand{\phan}{\ensuremath{\phantom{5}}}
\newcommand{\phal}{\phantom{<}}

\newcommand{\un}{}
\newcommand{\myem}{}

\newcommand{\rncnov}{\relax}
\newcommand{\ffsix}{\relax} 
\newcommand{\fmvsix}{\relax } 
\newcommand{\fmvrc}{\relax } 
\newcommand{\fmvcw}{\relax } 
\newcommand{\fmvrf}{\relax } 
\newcommand{\remark}[1]{\relax}

\long\def\inst#1{\par\nobreak\kern 4pt\nobreak
  {\it #1}\par\vskip 10pt plus 3pt minus 3pt}

\begin{document}

\begin{flushleft}
SLAC-PUB-\SLACPubNumber \\
\babar-PUB-\BABARPubYear/\BABARPubNumber \\
\end{flushleft}

\title{\large \bf
\boldmath
\Large
 Limits on the {\un {Decay-Rate}} Difference of Neutral-\B\ Mesons 
and on \CP, \T, and \CPT\ Violation in \BzBzb\ Oscillations
\begin{center} % Temporary fix for Collaboration name problem
\vskip 5mm
The \babar\ Collaboration
\end{center}
} % end title{}

\input pubboard/authors_jan2004.tex

\date{\today}

\begin{abstract}
\input abstract.tex
\end{abstract}

\pacs{13.25.Hw, 12.15.Hh, 14.40.Nd, 11.30.Er}% PACS, the Physics and Astronomy Classification Scheme.   \\

\maketitle

\newpage

\setcounter{footnote}{0}

\section{ \boldmath Introduction and analysis overview}
\label{sec:intro}
\input intro

\section{ \boldmath General time-dependent decay rates from $\Upsilon(4S)\to
\BzBzb$}
\label{sec:decayrates}
\input decayrates

\section{ \boldmath The \babar\ detector}
\label{sec:detector}
\input detector

\section{ \boldmath Data samples and \B-meson reconstruction}
\label{sec:sample}
\input sample

\section{ \boldmath Flavor tagging}
\label{sec:tagging}
\input tagging

\section{\boldmath Decay-time measurement and \dtresolutionfunction}
\label{sec:decaytime}
\input decaytime

\section{\boldmath Likelihood fit method}
\label{sec:method}
\input method

\section{ \boldmath Analysis results}
\label{sec:results}
\input results

\section{ \boldmath Cross-checks and validation studies}
\label{sec:checks}
\input checks

\section{ \boldmath Systematic uncertainties}
\label{sec:systematics}
\input systematics

\section{ \boldmath Summary and discussion of results}
\label{sec:summary}
\input summary

\section{ \boldmath Acknowledgments}
\label{sec:acknowl}
\input pubboard/acknowledgements

\appendix
\section{ \boldmath Efficiency asymmetries}
\label{appendix:asymmetries}
\input appendix

\input references
\end{document}

%% file: symbols.tex
\def\sintwob {\ensuremath{\sin{2\beta}}}
\def\Dstarp  {\ensuremath{D^{*+}}}
\def\Dstarm  {\ensuremath{D^{*-}}}

\def\Dstarz  {\ensuremath{D^{*0}}}
\def\Dp      {\ensuremath{D^+}}
\def\Dm      {\ensuremath{D^-}}

\def\Dz      {\ensuremath{D^0}}

%% file: symbolscpt.tex
\def\CPTv{\setlength{\unitlength}{8pt}\begin{picture}(3,1)\put(0.,0.){\CPT}\put(-0.1,-0.1){\line(3,1){2.6}}\end{picture}}
\def\CPv{\setlength{\unitlength}{8pt}\begin{picture}(2,1)\put(0.,0.){\CP}\put(-0.1,-0.1){\line(2,1){1.7}}\end{picture}}
\def\Tv{\setlength{\unitlength}{8pt}\begin{picture}(2,1)\put(0.,0.){~\T}\put(0.,0.){\line(2,1){1.6}}\end{picture}}

\def\CPscript  {\ensuremath{C\hspace{-0.15ex}\!P}\xspace}

\newcommand{\re}{\ensuremath{\mathop{\rm Re}}}
\newcommand{\im}{\ensuremath{\mathop{\rm Im}}}
\newcommand{\taub}{\ensuremath{\tau_B}}

\newcommand{\G}{\ensuremath{ \Gamma }}
\newcommand{\M}{\ensuremath{ m }}
\newcommand{\dMcpt}{\ensuremath{ \delta m }}
\newcommand{\dGcpt}{\ensuremath{ \delta \Gamma }}
\newcommand{\dGoverG}{\ensuremath{ \Delta \Gamma/\Gamma }}
\newcommand{\sgndGoverG}{\ensuremath{ \sign(\relambcpbare) \Delta \Gamma/\Gamma }}
\newcommand{\sgndGoverdM}{\ensuremath{ \sign(\relambcpbare) \Delta \Gamma/\dM }}
\newcommand{\dG}{\ensuremath{ \Delta \Gamma }}
\newcommand{\dM}{\ensuremath{ \Delta m }}

\newcommand{\absqop}{\ensuremath{|q/p|}}
\newcommand{\abspoq}{\ensuremath{|p/q|}}

\def\tagi{\ensuremath{{\rm tag}}}
\def\reci{\ensuremath{{\rm rec}}}
\def\flavi{\ensuremath{{\rm flav}}}
\def\tagbari{\ensuremath{{\overline{\rm tag}}}}

\def\tagip{{\rm ``tag''}}
\def\recip{{\rm ``rec''}}
\def\flavip{{\rm ``flav''}}
\def\CPip{\fmvrf{{\rm ``\CP''}}}
\def\tagbarip{\fmvrf{{``\ensuremath{\overline{\rm tag}}'' }}}

\def\tagv{\ensuremath{\B \hspace{-0.2ex}t }}
\def\flavv{\ensuremath{{\B \hspace{-0.4ex}f}}}
\def\tagbarv{\ensuremath{{\Bb \hspace{-0.2ex}t }}}
\def\flavbarv{\ensuremath{{\Bb \hspace{-0.4ex}f }}}

\def\lambrecbare {\ensuremath{ \lambda_{\reci} }}

\def\lambrecbarei {\lambrecbare}

\def\lambtagbare {\ensuremath{ \lambda_{\tagv} }}

\def\lambbartagbarbare {\ensuremath{ \overline{\lambda}_{\tagbarv }}}
\def\lambflavbare {\ensuremath{ \lambda_{\flavv} }}

\def\lambbarflavbarbare {\ensuremath{ \overline{\lambda}_{\flavbarv} }}

\def\imlambflavbare {\ensuremath{ \im \lambda_{\flavv} }}

\def\imlambbarflavbarbare {\ensuremath{ \im \overline{\lambda}_{\flavbarv} }}

\def\relambflavbare {\ensuremath{ \re \lambda_{\flavv} }}

\def\relambbarflavbarbare {\ensuremath{ \re \overline{\lambda}_{\flavbarv} }}
\def\lambflavbarbare {\ensuremath{ \lambda_{\flavbarv} }}

\def\imlambtagflat {\ensuremath{ \im \lambda_{\tagv} /|\lambda_{\tagv}| }}

\def\imlambbartagflat {\ensuremath{ \im \overline{\lambda}_{\tagv} /|\overline{\lambda}_{\tagv}| }}

\def\imlambflavflat {\ensuremath{ \im \lambda_{\flavv} /|\lambda_{\flavv}| }}

\def\imlambbarflavflat {\ensuremath{ \im \overline{\lambda}_{\flavv} /|\overline{\lambda}_{\flavv}|}}

\def\imlambcpflat {\ensuremath{ \im \lambda_{\CPscript} /|\lambda_{\CPscript}|}}

\def\lambcpbare {\ensuremath{ \lambda_{\CPscript} }}
\def\imlambcpbare {\ensuremath{ \im \lambda_{\CPscript}  }}
\def\relambcpbare {\ensuremath{ \re \lambda_{\CPscript}  }}

\def\matrixelement#1#2#3{\langle#1|#2|#3\rangle}

\def\resol{{\cal R}}

\def\hamiltonian{{\cal H}}

\newcommand{\ckm}{\ensuremath{ \rm CKM }}
\newcommand{\dckm}{\ensuremath{ \rm DCS }} %hcanged by rnc 05/27/03

\newcommand{\z}{\ensuremath{{\mathsf z}}\xspace}
\newcommand{\imZ}{\ensuremath{\im\z}}
\newcommand{\reZ}{\ensuremath{\re\z}}

\newcommand{\reZparflat}{\ensuremath{\left( \relambcpbare / |\lambcpbare| \right) \reZ }}

\newcommand{\res}{\ensuremath{\re s}}
\newcommand{\ims}{\ensuremath{\im s}}

\def\dt   {\ensuremath{\Delta t}}
\def\dtresolutionfunction{\dt\ resolution function}
\def\dtResolutionFunction{\dt\ Resolution Function}
\def\dz   {\ensuremath{\Delta z}}

\def\sdt{\sigma_{\dt}}

\def\dw   {\ensuremath{\Delta w}}

\def\dwa   {\ensuremath{\Delta w^\alpha}}
\def\wa   {\ensuremath{w^\alpha}}

\def\waj   {\ensuremath{w^{\alpha,j}}}
\def\dwasig   {\ensuremath{\Delta w^{\alpha,\rm sig}}}
\def\wasig   {\ensuremath{w^{\alpha,\rm sig}}}

\def\bflav   {\ensuremath{B_{\rm flav}}}
\def\bcpks   {\ensuremath{B_{{\CPscript}\KS}}}
\def\bcpkl   {\ensuremath{B_{{\CPscript}\KL}}}
\def\bcp   {\ensuremath{B_{\CPscript}}}
\def\btag   {\ensuremath{B_{\rm tag}}}
\def\brec   {\ensuremath{B_{\rm rec}}}

\def\mes   {\ensuremath{m_{\rm ES}}}
\def\de   {\ensuremath{\Delta E}}

\def\cpks   {\ensuremath{{\CP}}\KS}
\def\cpkl   {\ensuremath{{\CP}}\KL}

\def\sign{\mathop{\rm{sgn}}} 
\def\Rcp {\ensuremath{ r_{\CPscript} }} % changed by rnc 5/27/03

\def\palphar{\ensuremath{p^\alpha_{\reci}(\mes)}}

\def\nonleptonic{nonleptonic}
\def\rms{r.m.s.}

\def\mH{\ensuremath{m_H}}
\def\mL{\ensuremath{m_L}}
\def\gH{\ensuremath{\Gamma_H}}
\def\gL{\ensuremath{\Gamma_L}}
\def\mHL{\ensuremath{m_{H,L}}}
\def\gHL{\ensuremath{\Gamma_{H,L}}}

%% file: pubboard/authors_jan2004.tex
%% author list as of 02-Jan-2004 (592 authors)
%
\author{B.~Aubert}
\author{R.~Barate}
\author{D.~Boutigny}
\author{F.~Couderc}
\author{J.-M.~Gaillard}
\author{A.~Hicheur}
\author{Y.~Karyotakis}
\author{J.~P.~Lees}
\author{V.~Tisserand}
\author{A.~Zghiche}
\affiliation{Laboratoire de Physique des Particules, F-74941 Annecy-le-Vieux, France }
% ...
\collaboration{The \babar\ Collaboration}
\noaffiliation

%% file: abstract.tex
Using events in which one of two neutral-\B\ mesons from the decay of
an \FourS\ resonance is fully reconstructed, we set limits on
the difference between the {decay rates of the two} neutral-\B\ mass eigenstates
and on \CP, \T, and \CPT\ violation in $\BzBzb$ mixing.  
{The reconstructed decays, comprising both \CP\ and {flavor} eigenstates,} are obtained from 88 million
$\FourS \to \B\Bb$ decays collected 
with the \babar\ detector at 
the \pep2\ asymmetric-energy \B\ Factory at SLAC.  We determine
six independent parameters governing oscillations (\dM, \dGoverG), 
\CPT\ and \CP\ violation \mbox{(\reZ, \imZ)}, and \CP\ and \T\ violation
(\imlambcpbare, \absqop), where $\lambcpbare$ characterizes \Bz\ and \Bzb\ decays to
 states of charmonium plus \KS\ or \KL. The results are
$$\begin{array}{
r@{\ \ =\ }r@{.}l@{\pm 0.}l@{{\rm (stat.)}\pm 0.}
l@{{\rm{(syst.)}}\ \ [}r@{,\ }l@{]}l}
\sgndGoverG &-0&008&037&018&-0.084&0.068&~,\\
\absqop     &1 &029&013&011& 1.001&1.057&~,\\
\reZparflat &0 &014&035&034&-0.072&0.101&~,\\
\imZ        &0 &038&029&025&-0.028&0.104&~.
\end{array}$$
The values inside square brackets indicate the 90\% confidence-level
intervals. The values of \imlambcpbare\ and \dM\ are consistent
with previous analyses {\fmvcw and are used as cross-checks}.
These {\fmvsix measurements are in agreement} with
Standard Model expectations.

%% file: intro.tex
The mass difference \dM\ between the \Bz\ mass 
eigenstates has been measured 
with high precision at \B-factory experiments~\cite{ref:dM-babar-had,ref:dM-babar-dstlnu,ref:dM-babar-dilep,ref:dM-belle}, 
and \CP\ violation has been observed in neutral-\B-meson decays to 
states like $\jpsi \KS$~\mbox{\cite{ref:sin2b-babar,ref:sin2b-belle}}. However,
our knowledge of other aspects of {\fmvrc neutral}-\B-meson oscillations is meager. In this paper,
we provide direct limits on the total decay-rate difference \dG\ between
the \Bz\ mass eigenstates, and on \CP, \T, and \CPT\ violation due
to oscillations alone.

In the Standard Model, the ratio $\dG/\dM$ 
is of order $m_b^2/m_t^2$ and thus quite small. Recent calculations of
\dGoverG, including $1/m_b$ contributions and part of the
next-to-leading order QCD corrections \cite{ref:dighe, ref:ciuchini},
find values in the approximate range $-0.2\%$ to $-0.3\%$. 
Existing limits for $|\dGoverG|$~\cite{ref:cleochid,ref:delphi} are relatively weak ($\sim 20\%$).
The large data sets available {\fmvcw at asymmetric-energy} \B\ factories provide {\fmvcw an}
opportunity to look for deviations from the Standard Model.

The \CP-violating asymmetry observed in neutral-\B-meson decays to states like $\jpsi \KS$ 
is due to the interference between decay amplitudes to a \CP\ eigenstate with and without mixing.
\CP\ violation in mixing alone leads to different rates for the transitions \mbox{\Bz\to\Bzb} and \mbox{\Bzb\to\Bz.}
This can be measured, for example, by comparing
{the decay rates to \fmvrf{$\ell^-\ell^- X$} and \fmvrf{$\ell^+\ell^+ X$} from semileptonic
decays of pairs of neutral-\B\ mesons arising from the \FourS~\cite{ref:babardileptonTviolation}. 
{\fmvsix The only semileptonic decays generated by first-order weak interactions are
\fmvrf{$\Bz\to\ell^+\nu X$} and \fmvrf{$\Bzb\to\ell^-\overline{\nu} \overline{X}$} and the \CP\ invariance of strong
and electromagnetic interactions guarantees that these have equal rates. As a result, any asymmetry in the dilepton 
rates can be ascribed to \CP\ violation in mixing. While {\fmvrc \CP\ violation in mixing} is suppressed in the
Standard Model~\cite{ref:ciuchini,ref:absqopSM,ref:beneke},  
additional virtual contributions from new physics could obviate this suppression. 
Similarly, {\fmvrc new physics may introduce additional intrinsic
\T\ violation or even \CPT\ violation in mixing.} It is these possibilities for the breaking of
discrete symmetries in mixing itself that we address in this analysis
{\fmvsix using}
\nonleptonic\ decays that are completely reconstructed.}

The behavior of neutral-\B\ mesons is 
sensitive to \CPT\ violation~\cite{ref:sanda,ref:ko,ref:bbCPT}. 
A theorem~\cite{ref:cpttheo} founded on general
principles of relativistic quantum field theory
states that 
the \CPT\ symmetry holds for 
\fmvrf{any local field theory satisfying Lorentz invariance.}
The \CPT\ symmetry 
{\fmvrc is the only}
 combination of \C, \P, and \T\ that is not known to be violated.
Nevertheless, it is possible that \CPT\ symmetry could 
fail at short distances~\cite{ref:cptbreak}.
{\fmvrc Strict} constraints
on \CPT\ violation have been obtained in the neutral-kaon system~\cite{ref:cptkaons}.  
Limits in the \B-meson system have been obtained previously~\cite{ref:otherCPTBtests,ref:dM-belle}.

To measure  $\Delta\Gamma$ and  \CP, \T, or \CPT\ violation, we observe the time
dependence of decays of neutral-\B\ mesons produced in pairs at
the \FourS\ resonance.  The {\fmvrc usual approach to} mixing and 
\CP\ analyses~\cite{ref:dM-babar-had,ref:dM-babar-dstlnu,ref:dM-babar-dilep,ref:dM-belle,
ref:sin2b-babar,ref:sin2b-belle} allows 
for exponential decay, modulated by oscillatory terms with frequency \dM.
These analyses neglect the difference \dG\ between the decay rates
 of the two mass eigenstates, which would introduce
{\fmvrc terms with a new time dependence ${\un\exp(\pm\Delta\Gamma t/2)}$.} 
Violation of \CP, \T, or \CPT\  in the mixing of the
neutral-\B\ mesons would modify the coefficients of the various terms
involving exponential and oscillatory behavior.  
To detect these potential subtle changes requires precision measurements of the decays,
{\fmvrc detailed consideration of systematic effects, and thorough}
treatment of coherent production of neutral-\B-meson pairs from the $\Upsilon (4S)$.

This analysis is based on a total of about 88 million
$\FourS \to \B\Bb$ decays 
collected 
with the
\babar\ detector at the PEP-II asymmetric-energy
\B\ Factory at the Stanford Linear Accelerator Center. {\rncnov There, 9.0-\gev\ electrons} and 
3.1-\gev\ positrons annihilate to produce the $\B\Bb$ pairs moving
along the $e^-$ beam direction ($z$-axis) with a {\fmvcw Lorentz boost of
$\beta \gamma \approx 0.55$}. This boost makes it possible to measure
the proper-time difference \dt\ between the two \B\ decays.  
We fully reconstruct one meson from its decay to a flavor 
eigenstate (\bflav) or to a  \CP\ eigenstate  (\bcp) composed of charmonium and
either a \KS\ or \KL. We denote the flavor and \CP\ eigenstates jointly by \brec.
The remaining charged particles in the event, which originate from the
other \B\ meson (\btag), are used to identify (``tag'') its flavor as \Bz\ or \Bzb. 
Not all events can be tagged, but the untagged events are also used in 
the analysis.
The time difference $\dt \equiv t_{\rm rec} - t_{\rm tag} \approx
\Delta z / (\beta \gamma c)$ is determined from the separation 
{\fmvrc $\Delta z$ along the boost direction}
of the decay vertices for the fully reconstructed \B\ candidate and the tagging \B.

A {\fmvrc maximum-likelihood fit} to the time distributions of 
tagged and untagged, flavor and \CP\ eigenstates determines
six independent parameters (see Sec.~\ref{sec:decayrates}) governing oscillations (\dM, \dGoverG),
\CPT\ and \CP\ violation \mbox{(\reZ, \imZ)}, and \CP\ and \T\ violation (\imlambcpbare, \absqop), where 
$\lambcpbare$ is the usual variable used to characterize the decays of neutral-\B\ mesons into 
final states of charmonium and a \KS\ or \KL. The values of
\imlambcpbare\ and \dM\ are
used as cross-checks with the {\ffsix earlier} \babar\ $\sin2\beta$ result~\cite{ref:sin2b-babar}, 
{\fmvsix obtained with the same dataset}, and
with previous \B-factory measurements of \dM~\cite{ref:dM-babar-had,ref:dM-babar-dstlnu,ref:dM-babar-dilep,ref:dM-belle}. 
{\fmvcw All the parameters are explicitly defined in Sec.~\ref{sec:decayrates}}.

The analysis {\un presents} several challenges.  
First, the resolution for \dt\ is comparable to the
\B\ lifetime and is asymmetric in \dt.
This asymmetry must be well understood lest it be mistaken for a fundamental
asymmetry we seek to measure. 
 Second, tagging %
 assigns flavor incorrectly {some fraction of the time. 
Third, interference between 
{\fmvsix weak decays favored by the Cabibbo-Kobayashi-Maskawa (CKM) quark-mixing matrix and those
doubly-Cabibbo-suppressed (\dckm)}
cannot be neglected.
Fourth, direct \CP\ violation in the \bcp\ sample
could mimic \CP\ violation in mixing and must be parameterized
appropriately. Finally, we have to account for possible asymmetries
induced by the 
differing response of the detector to {\fmvrc positively and
negatively charged} particles. In resolving {\fmvrc all of the above}
issues we rely mainly on data.}

{\fmvsix This paper provides a detailed description of the analysis published in Ref.~\cite{ref:CPTprl},} 
and is organized as follows. In Sec.~\ref{sec:decayrates} we
present a general formulation of the time-dependent decay rates of
$\BzBzb$ pairs produced at the \FourS\ resonance, including
effects from the decay-rate difference, possible \CP\ and  \CPT\ violation in mixing, and
interference effects induced by \dckm\ decays. 
We derive the expressions for \B\ decays to flavor and \CP\ eigenstates. 
In Sec.~\ref{sec:detector} we
describe the \babar\ detector.  After discussing the data sample 
in Sec.~\ref{sec:sample}, we describe the \B-flavor tagging
algorithm in Sec.~\ref{sec:tagging}. Sec.~\ref{sec:decaytime} is devoted to the description of
the measurement of \deltaz and {to} the determination of \dt\ and its
resolution function. In Sec.~\ref{sec:method} we describe our
log-likelihood function and the assumptions made in the %nominal
fit. The results of the fit are given in Sec.~\ref{sec:results}.
Cross-checks are discussed in Sec.~\ref{sec:checks} and systematic
uncertainties are presented in Sec.~\ref{sec:systematics}. The
results of the analysis are summarized and discussed in
Sec.~\ref{sec:summary}.

%% file: decayrates.tex
The neutral-\B-meson system can be described by the effective Hamiltonian ${\bf H} = {\bf M}-i{\bf\Gamma}/2$,
where ${\bf M}$ and ${\bf\Gamma}$ are two-by-two Hermitian matrices
describing, respectively, the mass and decay-rate components.
{\fmvcw \CP\ or \CPT\ symmetry imposes that}   %!!!!!!!!!!!
$M_{11}=M_{22}$ and $\Gamma_{11}=
\Gamma_{22}$, the index $1$ indicating \Bz\  and $2$ indicating
\Bzb. 
{\fmvcw In the limit of \CP\ or \T\ invariance,}   %!!!!!!!!!!!!
$\Gamma_{12}/M_{12}=\Gamma_{21}/M_{21}=\Gamma_{12}^*/M_{12}^*$,
so $\Gamma_{12}/M_{12}$ is real. These conditions do not depend on the phase conventions 
chosen for the \Bz and \Bzb.  The masses $\mHL$ and decay rates $\gHL$ of the 
two eigenstates of ${\bf H}$  form the complex {\fmvrc eigenvalues} $\omega_{H,L}$
\begin{widetext}
\begin{equation}
\omega_{H,L} \equiv \mHL -\frac i2\gHL= \M-\frac i2 \G \pm
       \sqrt{\left(M_{12}-\frac i2\Gamma_{12}\right)\left(M_{12}^*-\frac i2\Gamma_{12}^*\right)+\frac 14\left(\dMcpt-\frac i2\dGcpt\right)^2}~,\label{eq:eigen-energies}
\end{equation}
\end{widetext}
{\rncnov where the real part of the square root is taken to be positive and} where we define
\begin{eqnarray}
\M \equiv \frac12 (M_{11}+M_{22}) & , &
\G \equiv \frac12 (\Gamma_{11}+\Gamma_{22}) ~, \nonumber \\
\dMcpt \equiv M_{11}-M_{22} \quad & , &
\dGcpt \equiv  \Gamma_{11}-\Gamma_{22}~.
\label{eq:defns}
\end{eqnarray}
Assuming \CPT\ invariance ($\dMcpt=0, \dGcpt=0$), and anticipating that  
$|\dG| \ll\dM$, we have
\begin{eqnarray}
 \dM & \equiv & \mH -\mL \approx 2|M_{12}| ~ , \nonumber\\
 \dG & \equiv &  \gH -\gL \approx 2|M_{12}| \re  (\Gamma_{12}/M_{12}) ~.
\end{eqnarray}
Here we have taken $\dM$ to be the mass of the heavier eigenstate minus
the mass of the lighter {\fmvsix one}. Thus, $\dG$ is the decay rate of the
heavier state minus the decay rate of the lighter {\fmvsix one} and its sign is not known a priori.

{\fmvcw With \CPT\ symmetry,}   %!!!!!!!!!!
the light and heavy mass eigenstates of the neutral-\B-meson system can
be written
\begin{eqnarray}
|B_L\rangle&=&p|\Bz\rangle +q|\Bzb\rangle\nonumber ~ ,\\
|B_H\rangle&=&p|\Bz\rangle -q|\Bzb\rangle ~ ,
\label{eq:mass_eigenstates}
\end{eqnarray}
where
\begin{equation}
 {q\over p}\equiv-\sqrt{{M^\ast_{12}-\frac{i}{2}\,\Gamma^\ast_{12}\over 
M_{12}-\frac{i}{2}\,\Gamma_{12}}} ~ .
\label{eq:qoverp}
\end{equation}
The magnitude of $q/p$  is very nearly unity:
\begin{equation}
\left|\frac qp\right|^2\approx 1-\im \frac{\Gamma_{12}}{M_{12}}~.
\end{equation}

In the Standard Model, the \CP- and \T-violating quantity $|q/p|^2-1$ is
small not just because $|\Gamma_{12}|$ is small, but additionally
because the \CP-violating quantity $\im (\Gamma_{12}/M_{12})$
{\fmvcw is suppressed by an additional factor
$(m_c^2-m_u^2)/m_b^2\approx 0.1$ relative to $|\Gamma_{12}/M_{12}|$.}
Violation of \CP\ is not possible if two of the quark masses (for quarks
of the same charge) are
identical, for then we could redefine 
two new quark states with equal masses so that one of them did not mix with the two remaining states.
The mixing among two generations would be inadequate to support \CP\ violation. 
When the remaining Standard Model factors are included, the expectation is 
\mbox{$|\im ({\Gamma_{12}}/{M_{12}})|<10^{-3}$}~\cite{ref:ciuchini,ref:absqopSM,ref:beneke}.
 
\CPT\ violation in mixing can be described conveniently
by the 
phase-convention--independent 
quantity
\begin{eqnarray}
\z &\equiv & {\dMcpt-\frac i2\,\dGcpt\over
2\sqrt{\left(M_{12}-\frac{i}{2}\,\Gamma_{12}\right)
\left(M^\ast_{12}-\frac{i}{2}\,\G^\ast_{12}\right)
+ \frac14\left(\dMcpt - \frac{i}{2}\,\dGcpt\right)^2}} \nonumber \\ 
& {\un =} & \frac{\dMcpt-\frac i2\,\dGcpt}{  \dM - \frac i2 \dG}~.\label{eq:zdefn}
\end{eqnarray}
The generalization of the eigenstates in Eq.~(\ref{eq:mass_eigenstates}) when we account for
\CPT\ violation can be written
\begin{eqnarray}
|B_L\rangle&=&p\sqrt{1-\z}|\Bz\rangle +q\sqrt{1+\z}|\Bzb\rangle\nonumber ~ ,\\
|B_H\rangle&=&p\sqrt{1+\z}|\Bz\rangle -q\sqrt{1-\z}|\Bzb\rangle ~ ,
\label{eq:mass_eigenstates_cpt}
\end{eqnarray}
where we maintain the definition of $q/p$ given in Eq.~(\ref{eq:qoverp}).
The result, when time evolution is included, is that
states that begin as purely \Bz\ or \Bzb\ after a time $t$ will be mixtures
\begin{eqnarray}
\ket{{\Bz}_{\text{phys}}(t)} &=& 
\bigl[ g_+(t) + \z g_-(t)\bigr]\,\ket{\Bz} \nonumber \\
& & - \sqrt{1-\z^2} \frac{q}{p}\,g_-(t)\,\ket{\Bzb}~, \\
\ket{{\Bzb}_{\text{phys}}(t)} &=&
\bigl[g_+(t) - \z g_-(t)\bigr]\,\ket{\Bzb} \nonumber \\
& & - \sqrt{1-\z^2} \frac{p}{q}\,g_-(t)\,\ket{\Bz}~,
\label{eq:general}
\end{eqnarray}
where we have introduced
\begin{equation}
g_\pm(t)=\frac 12(e^{-i\omega_Ht}\pm e^{-i\omega_Lt})~.\label{eq:gdefn}
\end{equation}
Invariance under \CP\ or under \T\ requires that 
\begin{equation}
|\langle \Bz\ket{{\Bzb}_{\text{phys}}(t)}|=
 |\langle \Bzb\ket{{\Bz}_{\text{phys}}(t)}|~;
\end{equation} i.e., 
$|q/p|=1$, which is guaranteed by $\im(\Gamma_{12}/M_{12})=0$.
{\fmvsix Table \ref{tab:cpt} shows the constraints on \absqop\ and \z\ for the
different possible symmetry scenarios. The Standard Model corresponds to the 
second configuration (\CPT\ {\fmvcw symmetry}, with \CP\ and \T\ violated).  %!!!!!!!!!!
Note that two of these scenarios are degenerate. 
{\fmvcw With \CP\ symmetry in \BzBzb\ oscillations,}    %!!!!!!!!!
this experiment cannot distinguish between \T\ and \CPT\ both being conserved or violated.}

\def\equalZero{{$=0$}}
\def\notequalZero{{$\neq 0$}}
\def\equalOne{{$=1$}}
\def\notequalOne{{$\neq 1$}}

\begin{table}[htb]
\centering
\caption{{\fmvrc Constraints on} \absqop\ and \z\ due to \CP, \T, and \CPT\ symmetries in \BzBzb\ oscillations. 
}
\label{tab:cpt}
 \begin{tabular}{rl|c|c|c|c|c}
   & & \multicolumn{2}{c|}{\CPT} & \multicolumn{3}{c}{\CPTv} \\
 \cline{3-7}
   & & ~\CP,~\T~ & ~\CPv,~\Tv & ~\CPv,~\T~ & ~\CP,~\Tv   & ~\CPv,~\Tv \\
 \hline \hline
    & & & & & & \\
 {\mbox{$\absqop$}} &
 %$M_{12}$, $\Gamma_{12}$ relatively real &
   &
 {\equalOne} & {\notequalOne} & {\equalOne} & {\equalOne} & {\notequalOne} \\
     &  & & & & & \\
\hline
    & & & & & & \\
 {$\z$} &    &
 {\equalZero} & {\equalZero} & {\notequalZero} & {\equalZero} & {\notequalZero} \\
    & & & & & & \\
 \hline
\end{tabular}
\end{table}

\subsection{Effects of Coherence}
At the $\Upsilon(4S)$ resonance, neutral-\B\ mesons are produced in coherent 
{p-wave} pairs.  
If we subsequently {\rncnov observe one \B-meson decay to the state $f_1$ at time
$t_0=0$ and the other decay to the state $f_2$ at some later time $ t$, 
we cannot in general know whether $f_1$ came from the
decay of a \Bz\ or a \Bzb, and similarly for the state $f_2$.}  If
$A_{1,2}$ and ${\overline A}_{1,2}$ are the amplitudes for the decay
of \Bz\ and \Bzb, respectively, to the states $f_1$ and $f_2$, then the overall amplitude 
is given by 
\begin{eqnarray}
{\cal A}
&=&a_+g_+(t)+a_-g_-(t)~,
\end{eqnarray}
where
\begin{eqnarray}
a_+&=&-A_{1}{\overline A}_2+{\overline A}_{1} A_2\nonumber ~, \\
a_-&=&\sqrt{1-\z^2}\left[{p\over q}A_{1}A_2-{q\over p}{\overline A}_{1}{\overline A}_2\right] + \z\left[ A_{1} {\overline A}_2+{\overline A}_{1}A_2 \right]~.\nonumber \label{eq:apm}\\
\end{eqnarray}

Using the relations \begin{equation}
|g_\pm(t)|^2 = \frac{1}{2}\,e^{-\G t}\left[
\cosh(\dG t/2) \pm \cos(\dM t)\right]
\label{eq:gpmdefn}
\end{equation}
and
\begin{equation}
g_+^\ast(t)\,g_-(t) = -\frac{1}{2}\,e^{-\G t}\left[
\sinh(\dG t/2) + i\,\sin(\dM t)\right]~,
\label{eq:gpgmdefn}
\end{equation}
we find the decay rate
\begin{eqnarray}
{{\rm d}N\over {\rm d}t} &\propto & e^{-\G |t|}\Biggl\{ {1\over 2} c_+\cosh(\dG t/2)
+{1\over 2}c_-\cos(\dM t) \nonumber \\
& & \qquad -\res\, \sinh(\dG t/2)
+\ims\,\sin(\dM t) \Biggr\}~, ~~~~~~\label{eq:pref2}
\end{eqnarray}
where
\begin{equation}
c_\pm=|a_+|^2\pm|a_-|^2~,\quad s=a_+^*a_-~.\label{eq:cpms}
\end{equation}
The absolute value in the leading exponential in Eq.~(\ref{eq:pref2}) is introduced for
later convenience.

Now let us take $f_1 \equiv f_{\tagi}$ to be the state that is
incompletely reconstructed and that provides the tagging decay, and
$f_2 \equiv f_{\reci}$ to be the fully reconstructed state (flavor or \CP\ eigenstate).  
Then {\fmvsix we have} $t=t_{\reci}-t_{\tagi}$ and Eq.~(\ref{eq:apm}) becomes

\begin{eqnarray}
a_+&=&-A_{\tagi}{\overline A}_\reci+{\overline A}_{\tagi} A_\reci\nonumber ~, \\
a_-&=&\sqrt{1-\z^2}\left[{p\over q}A_{\tagi}A_\reci-{q\over p}{\overline A}_{\tagi}{\overline A}_\reci\right] \nonumber\\
&&+ \z\left[ A_{\tagi} {\overline A}_\reci+{\overline A}_{\tagi}A_\reci \right]~.\label{eq:apm_tag_reco}
\end{eqnarray}

If instead the tagged
decay occurs second, we would need to redefine $t$, $a_+$ and $a_-$ by interchanging 
the labels \tagip\ and \recip. This would amount to the replacements $t\to -t$, $a_+\to-a_+$, and 
$a_-\to a_-$.  However, we see that Eq.~(\ref{eq:pref2}) is actually unaffected by these 
changes and that we can instead retain the definitions $t=t_{\reci}-t_{\tagi}$ and those 
of Eq.~(\ref{eq:apm_tag_reco}). Thus, Eqs.~(\ref{eq:pref2})-(\ref{eq:apm_tag_reco}) apply independent 
of the order of the decays of the tagged and fully reconstructed \B mesons.

{\fmvrc A fully reconstructed flavor state cannot always be}
unambiguously associated with either \Bz\ or \Bzb. 
\dckm\ decays, such as $\Bz\to D^+\pi^-$, occur
at a rate suppressed by roughly $|V_{ub}^*V_{cd}^{}/V_{cb}^*V_{ud}^{}|^2 \approx (0.02)^2$. 
{\fmvrc Although this can be neglected,}
interference between favored
and suppressed {\ffsix amplitudes}
is reduced by only a factor of
approximately $0.02$~\cite{ref:owenandus}, {\fmvsix and must be taken into account.}  

Tagging cannot be done perfectly, largely because the tagging state is
incompletely reconstructed.  We account for this by measuring 
the wrong-tag probability {\ffsix from the data}.  However, even if our tagging were perfect in
principle, it would be afflicted with the same {\fmvrc complication from \dckm\ decays} as the fully
reconstructed state.
The full expressions for the real
coefficients $c_\pm$ and the complex coefficient $s$, 
{\fmvcw containing}
the \dckm\ amplitudes, are

\newcommand{\Ar}{\ensuremath{A_{\reci}}}
\newcommand{\At}{\ensuremath{A_{\tagi}}}
\newcommand{\Arb}{\ensuremath{\overline{A}_{\reci}}}
\newcommand{\Atb}{\ensuremath{\overline{A}_{\tagi}}}
\newcommand{\qop}{\ensuremath{\frac{q}{p}}}
\newcommand{\poq}{\ensuremath{\frac{p}{q}}}
\newcommand{\qopflat}{\ensuremath{q/p}}
\newcommand{\poqflat}{\ensuremath{p/q}}

\begin{widetext}

\begin{eqnarray}
c_{\pm} &=& \biggl\{
\left|\Arb\At - \Ar\Atb\right|^2
\pm |\z|^2 \left|\Arb\At + \Ar\Atb\right|^2
\pm |1-\z^2|\,\left|\poq\Ar\At - \qop\Arb\Atb\right|^2 \nonumber \\
&& \quad
\pm 2{\rm Re}\left[
\z^{\ast}\sqrt{1-\z^2}\,
\left(\poq\Ar\At - \qop\Arb\Atb\right)
\left(\Arb\At + \Ar\Atb\right)^{\ast}
\right]
\biggr\} 
\label{eq:cplusminus} \\
s &=& \biggl\{
\left(\Ar\Atb - \Arb\At\right)^{\ast}
\left[
\sqrt{1-\z^2}\left(\poq\Ar\At - \qop\Arb\Atb\right)
+ \z\left(\Arb\At + \Ar\Atb\right)
\right]
\biggr\}~.
\label{eq:s}
\end{eqnarray}

\input coeff-tables

\end{widetext}
Terms proportional {\fmvrc to} \Ar\Atb\ and \Arb\At\ are associated with decays
with no net oscillation between the two neutral-\B\ decays, while terms
proportional to (\qopflat)\Arb\Atb\ and (\poqflat)\Ar\At\ represent a net oscillation.

We characterize each final state $f$ through the parameter
\begin{equation}
\lambda_{{f}} = \frac qp \frac{\overline A_f}{A_f}\label{eq:lambdaf}~,
\end{equation}
where $f$ can {\fmvrc be \recip (which can be itself \flavip\ or \CPip) or \tagip.} 
In the absence of \dckm\ decays, $\lambda_{\flavi}$ (i.e., $\lambda_{\reci}$ when
the reconstructed state is a flavor eigenstate, not a \CP\ eigenstate)
and $\lambda_{\tagi}$ would be either zero or infinite.  With a {\fmvrc contribution from \dckm\ decays}
they are non-zero and finite.

If the reconstructed flavor state $f_{\flavi}$ is 
ostensibly a \Bz\ (hereafter indicated 
as $\flavv$ %\remarkrc{note notational change} 
to avoid ambiguities with the tag state) then
$|\lambrecbarei| \equiv |\lambflavbare| \ll 1$.
Conversely, if the
reconstructed state appears to come from a
\Bzb\ (indicated as $\flavbarv$), then $|\lambflavbarbare|\gg 1$, and it is convenient to
introduce $\lambbarflavbarbare \equiv 1/\lambflavbarbare$.  The pattern for the tagging state (\tagip$=\tagv,\tagbarv$) 
is similar.  If the reconstructed state is a 
\CP\ eigenstate, then $|\lambrecbare|\equiv |\lambcpbare|$ is of order
unity.

In practice, terms quadratic in $\z$ or in a small $\lambda_f$ are not
important.  The expressions for $c_\pm$ and $s$ when only linear terms
in small quantities are retained are shown in
Tables~\ref{table:cplus}, ~\ref{table:cminus}, and \ref{table:s}.  
The analysis uses the full expressions, without simplification.

It is appropriate to assume that the decays to flavor 
eigenstates {\rncnov we consider are dominated by a single weak mechanism: $\b\to \c \ubar \d$.   
{\fmvsix While we can find a mechanism for $\bbar\to \c \ubar \d$ (which is a \dckm\ process), 
there are no alternative first-order weak processes that produce $\c\ubar\d$ from a $\b$ quark.}
Then even if there are several contributions to the decay, each possibly with its own strong phase, 
the \CP-conjugate decay differs only
by changing a single common weak phase}  so that
$|A_{\flavv}|=|{\overline A}_{\flavbarv}|$, 
$|{\overline{A}}_{\flavv}|=|{A}_{\flavbarv}|$ {\fmvsix (and similarly for tagging states)}.
{In fact, even if this assumption is not rigorously true, any violation will be absorbed in tagging and reconstruction efficiencies, which are determined
from the data, as described in Sec.~\ref{sec:method}.}
{ These equalities} relate the four permutations that arise from the tag and reconstructed state being 
either \mbox{\Bz\ or \Bzb.}

\subsection{Ensembles of States}
\def\had {\ensuremath{ h }}
\def\hadbar {\ensuremath{ \overline{h} }}
\def\fhad {\ensuremath{ f_{{ h}} }}
\def\fbarhad {\ensuremath{ \overline{f}_{{ h}} }}
In principle, {\fmvrc every hadronic final
state \fhad} has a different $\lambda_{\had}$,%\remarkrc{note notational change} 
which can be {\fmvsix written as
$\lambda_{\had}=|\lambda_{\had} |e^{-i\phi_{\had}^\mathrm{even}}e^{-i\phi_{\had}^\mathrm{odd}}$, where
$\phi_{\had}^\mathrm{even}$ and $\phi_{\had}^\mathrm{odd}$ are strong (\CP-even) and weak (\CP-odd) phases that arise 
from the ratio of the amplitudes of the \Bz\ and \Bzb\ decays to \fhad.
Assuming that there is a single weak phase involved,
the \CP-conjugate state \fbarhad\ {\fmvrc will have} 
${\overline\lambda}_{\hadbar}=\abspoq^2 |\lambda_{\had} 
|e^{-i\phi_{\had}^\mathrm{even}}e^{i\phi_{\had}^\mathrm{odd}}$.}

\def\Fcoll {\ensuremath{ F }}
\def\Fcollbar {\ensuremath{ F }}
\def\Tcoll {\ensuremath{ \alpha }}
\def\Tcollbar {\ensuremath{ \alpha }}

{If we sum {\rncnov squares of amplitudes} over a collection \Fcoll\ of flavor states that are ostensibly
\Bz, the terms that do and do not contain $\lambda_{\flavi}$ are of the form
}
\begin{equation}
\sum_{\fhad\in \Fcoll}|A_{\had}|^2\lambda_{\had} \qquad{\rm and}\qquad
\sum_{\fhad\in \Fcoll}|A_{\had}|^2 ~,
\end{equation}
so we can define an effective $\lambda_{\flavv}$ by
\begin{equation}
\lambda_{{\flavv}}=\frac{\sum_{\fhad\in \Fcoll}|A_{\had}|^2\lambda_{\had}}{\sum_{\fhad\in \Fcoll}|A_{\had}|^2} ~.
\end{equation}
Similarly, for flavor states that {\fmvrc are ostensibly \Bzb}, 
\begin{equation}
{\overline \lambda}_{\flavbarv}=\frac{\sum_{\fbarhad\in\Fcollbar}|\overline{A}_{\hadbar}|^2
{\overline{\lambda}}_{\hadbar}}{\sum_{\fbarhad\in\Fcollbar}|\overline{A}_{\hadbar}|^2} ~.
\end{equation}
{\un The two complex numbers $\lambda_{\flavv}$ and ${\overline \lambda}_{\flavbarv}$ 
encapsulate the {\fmvrc effects due to \dckm\ decays} in the fully reconstructed \B\ decay, as long as 
the terms quadratic 
in $\lambda_{\had}$ and $\overline{\lambda}_{\hadbar}$, 
suppressed by roughly $|V_{ub}^*V_{cd}^{}/V_{cb}^*V_{ud}^{}|^2 \approx (0.02)^2$,
are omitted.

The same argument applies to tagging states. If the collection of states {\fmvrc contributing to 
a \Bz\ or \Bzb\ tag 
is \Tcoll\ then} 
\begin{eqnarray}
\lambda_{{\tagv}}&=&\frac{\sum_{\fhad\in \Tcoll}|A_{\had}|^2\lambda_{\had}}{\sum_{\fhad\in \Tcoll}|A_{\had}|^2} ~,
\\
{\overline \lambda}_{\tagbarv}&=&\frac{\sum_{\fbarhad\in\Tcollbar}|\overline{A}_{\hadbar}|^2
\overline{\lambda}_{\hadbar}}{\sum_{\fbarhad\in\Tcollbar}|\overline{A}_{\hadbar}|^2} ~.
\end{eqnarray}
In practice, we do not use separate $\lambda_{\tagi}$ parameters for each tagging
category \Tcoll\ {\fmvrc (i.e., each collection of states of similar character, 
as described in Sec.~\ref{sec:tagging})}, 
but simply one for \Bz\ and one for \Bzb, setting aside the
lepton tag category, 
{\fmvrc which is free of \dckm\ decays}.
{\fmvsix This treatment is flexible enough to incorporate the {\fmvrc \dckm-decay effects} that 
can mimic the {\fmvrc asymmetries} we seek in the analysis.}

Henceforth, expressions like
$\lambda_{\flavi}$ and $\lambda_{\tagi}$ refer to an appropriate sum over observed
states. The summation over states \fhad\ in a tagging category should be thought of as 
extending over those states that are reconstructed  
as belonging to {\fmvrc the given} category. In
this way, we incorporate implicitly the tagging efficiency of each state \fhad. 
The reconstruction efficiency is incorporated in an analogous fashion into $\lambda_{\flavi}$.

Data from directly related \CP\ final states like $\jpsi\KS$, with $\eta_{\CPscript}=-1$, 
and $\jpsi\KL$, with $\eta_{\CPscript}=+1$, 
where $\eta_{\CPscript}$ is the \CP\ eigenvalue of the final state,
can be combined by assuming that their time distributions are identical, except
for the factor $\eta_{\CPscript}$. % multiplying $\lambcpbare$.
We use a single parameter \lambcpbare\ obtained multiplying Eq.~(\ref{eq:lambdaf}) by $\eta_{\CPscript}$.
We assume $\Rcp=|\overline A_{\CPscript}/A_{\CPscript}|=1$
as expected theoretically at the
$10^{-3}$ level~\cite{ref:grossman} and as supported {\fmvrc experimentally by}: 
\begin{enumerate}[topsep=1pt, partopsep=1pt, parsep=1pt, itemsep=1pt, label=\emph{\roman*})]
\item the average of \B-factory measurements
of states of charmonium and \KS\ or \KL, from which it has 
been obtained $r_{\CPscript}=0.949\pm0.045$~\cite{ref:sin2b-babar,ref:sin2b-belle},
when \dG, \mbox{$\absqop-1$} and \z\ are assumed to be zero; 
\item the average of CLEO and \babar\ measurements of 
the \CP\ asymmetry in the charged 
mode \mbox{$\B^\pm \rightarrow \jpsi K^\pm$}, from which it is 
found $r_{\CPscript,\jpsi K^\pm}=1.008\pm0.025$~\cite{ref:cleojpsikp,ref:babarjpsikp}, combined
with isospin symmetry to relate with the \CP\ final states~\cite{ref:theojpsikp}.
\end{enumerate}

\subsection{Sensitivity of Distributions to Parameters}

{\fmvsix From Eq. (\ref{eq:pref2}) and
{\un\myem Tables~\ref{table:cplus}, \ref{table:cminus}, and
\ref{table:s}, it can be seen that} 
{\rncnov while \imlambcpbare, \imZ, \absqop, and \Rcp\ are unambiguously 
determined, \reZ\ appears only in the product
\relambcpbare\reZ\ or else is  suppressed by the small factor $\dG/\G$.  Similarly, the sign of \dG\ cannot be
determined separately from the sign of \relambcpbare\ since \dG\ always
appears multiplied by  \relambcpbare\ in its
dominant contribution.} Its value is known only 
through $\relambcpbare=\pm\sqrt{|\lambcpbare|^2-(\imlambcpbare)^2}$, where the choice of sign
could be made by a {\fmvrc separate}
measurement that directly determines the sign of \relambcpbare.
As a result, the parameters that {\fmvrc can be} determined by
this analysis are \sgndGoverG, \absqop, \reZparflat, \imZ,
\imlambcpflat, \Rcp, \dM,  and $\G$.  
In practice, we fix \Rcp\ and \G\ in the nominal fit, and vary them for systematic
studies.

Data for {\fmvrc final} states that are \CP\ eigenstates and those that are flavor eigenstates
are {\fmvrc both} needed for the analysis, as shown in Table~\ref{table:sensitivity}.
The sensitivity to \reZparflat\ and \imlambcpflat\ is provided by the
{\fmvsix decays to \CP\ eigenstates \bcp,}
for which the {\un accompanying} $t$ dependence is even for
the former and odd for the latter. The \bflav\ sample contributes
marginally {\fmvcw to} these parameters because it lacks explicit dependence on \imlambcpflat\ and
the dependence on \reZ\ is scaled by the $\sinh\left( \dG t / 2
\right)$ term, which is small for small \dG. 

In contrast, {\fmvrc the parameters} $\absqop$ and
\imZ\ (and \dM) are {\fmvrc determined} by the large \bflav\ sample, {\un where the
former is associated with a $t$-even distribution and the latter with a $t$-odd distribution.} For small values of \dGoverG, the determination of \dGoverG\
is dominated by the \bcp\ sample, {\un despite the smallness
of this sample compared to the \bflav\ sample}. 
This is {\fmvrc because in the} flavor
sample the leading dependence on \dG\ is {\fmvrc proportional to} $\dG^2$, while in the \CP\ {\fmvrc sample 
it is proportional to} \dG.
The contribution of $\sinh(\dG
t/2)$ is the same for both \Bz\ and \Bzb\ tags, so events {\fmvrc that cannot be tagged}
may be included {\fmvrc in the analysis to improve sensitivity}. The \bcp\ sample is also sensitive to the sign of
\dGoverG\ (up to the sign ambiguity from \relambcpbare). 

Overall, the combined use of the
\bflav\ and \bcp\ samples provides sensitivity to the full set of physical
parameters, since they are determined either from different samples,
or from different $t$ dependences.

\begin{table}
\centering
\caption[a]{\fmvrf{Dominant dependence} of the time distributions on the physical 
parameters
measured {\fmvcw with fully} reconstructed flavor and \CP\ states. Sensitivity is
specific to terms in the time dependence that are either $t$-even or $t$-odd. The flavor sample
is much larger than the \CP\ sample.}\label{table:sensitivity}
\vspace{0.1in}
\setlength{\extrarowheight}{3pt}

\begin{tabular*}{0.95\columnwidth}%
     {@{\extracolsep{\fill}}l|c|c|c|c}
\hline
&\multicolumn{2}{c|}{\bflav}&\multicolumn{2}{c}{\bcp}\\ 
Parameter & $t$-even~~ & $t$-odd~~ & $t$-even~~ & $t$-odd~~ \\ \hline\hline
$\absqop$& $\times$~ &&&\\ 
\dM & $\times$~ &&&\\ \hline
\imZ&& $\times$~ &&\\ \hline
$\reZparflat$&&& $\times$~ & \\ 
$\Rcp$&&& $\times$~ &\\ \hline
$\sgndGoverG$&&&& $\times$~ \\ 
$\imlambcpflat$&&&& $\times$~ \\ \hline
\end{tabular*}
\end{table}

{\fmvrc As we show in} 
Tables~\ref{table:cplus}, \ref{table:cminus}, and \ref{table:s}, 
if the reconstructed state is a flavor eigenstate, the {\fmvrc \dckm-decay effects} 
in tagging are negligible except in the $\sin(\dM t)$
term, {for the other terms are suppressed by both a power of $\lambda_{\flavi}$ and
a power of $\lambda_{\tagi}$.
Conversely, if the reconstructed state is a \CP\ eigenstate with
$|\lambcpbare| \approx 1$, the {\fmvrc effects from \dckm\ decays are} confined to the terms even \mbox{in $t$.}

%% file: coeff-tables.tex
\begin{table}[ht]
\centering
\caption{\label{table:cplus}The coefficient $c_+$ from Eq.~(\ref{eq:cplusminus}), evaluated to leading
order in the small quantities $\z$, $\lambflavbare$, $\lambtagbare$,
$\lambbarflavbarbare$, and $\lambbartagbarbare$.  If the tagging state for
a \Bz\ tag is $f_{\tagv}$, then the tagging state for a \Bzb\ is the
\CP-conjugate state, $f_{\tagbarv}$, and similarly for the fully
reconstructed
states. The decay amplitudes are $A_{\tagv}=\matrixelement{f_{{\tagv}}}{\hamiltonian}{\Bz}$,  ${\overline A}_{{\tagv}}=\matrixelement{f_{{\tagv}}}{\hamiltonian}{\Bzb}$,
$A_{\tagbarv}=\matrixelement{f_{\tagbarv}}{\hamiltonian}{\Bz}$,  ${\overline A}_{\tagbarv}=\matrixelement{f_{\tagbarv}}{\hamiltonian}{\Bzb}$,
and similarly for $\reci = \flavv, \flavbarv, \CP$.}
\setlength{\extrarowheight}{0.1in}
\begin{tabular}{ll|l}\hline
\btag&\brec& $c_+$\\[0.05in] \hline\hline
\Bz\quad&\Bz\quad&$|A_{\tagv}|^2|A_{\flavv}|^2\,|p/q|^2$\\
\Bz\quad&\Bzb\quad&$|A_{\tagv}|^2|\overline A_{\flavbarv}|^2$\\
\Bzb\quad&\Bz\quad&$|\overline A_{\tagbarv}|^2|A_{\flavv}|^2$\\
\Bzb\quad&\Bzb\quad&$|\overline A_{\tagbarv}|^2|\overline A_{\flavbarv}|^2\,|q/p|^2$\\
\Bz\quad&$B_{\CPscript}$\quad&$|A_{\tagv}|^2|A_{\CPscript}|^2\,|p/q|^2[
1+|\lambcpbare|^2 -4\re\lambcpbare\re\lambtagbare +2\re \z \re \lambcpbare-2\im \z \im \lambcpbare]$\\
\Bzb\quad&$B_{\CPscript}$\quad&$|\overline{A}_{\tagbarv}|^2|A_{\CPscript}|^2\,[
1+|\lambcpbare|^2 -4\re\lambcpbare\re\lambbartagbarbare -2\re \z \re \lambcpbare -2\im \z \im \lambcpbare]$\\[0.05in] \hline
\end{tabular}
\end{table}
\begin{table}[ht]
\centering
\caption{\label{table:cminus}The coefficient $c_-$ from Eq.~(\ref{eq:cplusminus}), evaluated to leading
order in the small quantities $\z$, $\lambflavbare$, $\lambtagbare$,
$\lambbarflavbarbare$, and $\lambbartagbarbare$. See caption of Table~\ref{table:cplus} for the definition of the 
various quantities.}
\setlength{\extrarowheight}{0.1in}
\begin{tabular}{ll|l}\hline
\btag&\brec& $c_-$\\[0.05in] \hline\hline
\Bz\quad&\Bz\quad&$-|A_{\tagv}|^2|A_{\flavv}|^2\,|p/q|^2$\\
\Bz\quad&\Bzb\quad&$|A_{\tagv}|^2|\overline{A}_{\flavbarv}|^2$\\
\Bzb\quad&\Bz\quad&$|\overline A_{\tagbarv}|^2|A_{\flavv}|^2$\\
\Bzb\quad&\Bzb\quad&$-|\overline A_{\tagbarv}|^2|\overline A_{\flavbarv}|^2\,|q/p|^2$\\
\Bz\quad&$B_{\CPscript}$\quad&$|A_{\tagv}|^2 |A_{\CPscript}|^2\,|p/q|^2[
-1+|\lambcpbare|^2 -4\im\lambcpbare\im\lambtagbare -2\re \z \re \lambcpbare +2\im \z \im \lambcpbare]$\\
\Bzb\quad&$B_{\CPscript}$\quad&$|\overline A_{\tagbarv}|^2|A_{\CPscript}|^2\,[
1-|\lambcpbare|^2 +4\im\lambcpbare\im\lambbartagbarbare +2\re \z \re \lambcpbare+2\im \z \im \lambcpbare]$\\[0.05in] \hline
\end{tabular}
\end{table}
\begin{table}[ht]
\centering
\caption{\label{table:s}The complex coefficient $s$ from Eq.~(\ref{eq:s}), 
evaluated to leading
order in the small quantities $\z$, $\lambflavbare$, $\lambtagbare$,
$\lambbarflavbarbare$, and $\lambbartagbarbare$. See caption of Table~\ref{table:cplus} for the definition of the 
various quantities.}
\setlength{\extrarowheight}{0.1in}
\begin{tabular}{ll|l}\hline
\btag&\brec& $s$\\[0.05in] \hline\hline
\Bz\quad&\Bz\quad&\fmvrf{$|A_{\tagv}|^2|A_{\flavv}|^2\,|p/q|^2[\lambtagbare^* -\lambflavbare^* ]$}\\
\Bz\quad&\Bzb\quad&$|A_{\tagv}|^2|\overline A_{\flavbarv}|^2
[\lambtagbare-\lambbarflavbarbare-\z]$\\
\Bzb\quad&\Bz\quad&$|\overline A_{\tagbarv}|^2|A_{\flavv}|^2
[\lambbartagbarbare-\lambflavbare+\z]$\\
\Bzb\quad&\Bzb\quad&$|\overline A_{\tagbarv}|^2|\overline A_{\flavbarv}|^2\,|q/p|^2
[\lambbartagbarbare^*-\lambbarflavbarbare^*]$\\
\Bz\quad&$B_{\CPscript}$\quad&\fmvrf{$|A_{\tagv}|^2|A_{\CPscript}|^2\,|p/q|^2[
|\lambcpbare|^2\lambtagbare  - \lambcpbare^* + \lambtagbare^* - |\lambcpbare|^2 \z ]$}\\
\Bzb\quad&$B_{\CPscript}$\quad&\fmvrf{$|\overline{A}_{\tagbarv}|^2|A_{\CPscript}|^2\,[
\lambbartagbarbare - \lambcpbare  + |\lambcpbare|^2\lambbartagbarbare^* + \z ]$}\\[0.05in] \hline
\end{tabular}
\end{table}

%% file: detector.tex
The \babar\ detector is described in detail
elsewhere~\cite{ref:babar-nim}, so here we give only a brief
description of the apparatus.

Surrounding the beam-pipe is a five-layer silicon vertex tracker (SVT),
which gives precisely measured points along the trajectories of charged
particles as they leave the interaction region.
Outside the SVT is a 40-layer drift chamber (DCH) filled with an 80:20
helium-isobutane gas mixture, chosen to minimize multiple scattering.
Charged-particle tracking and the determination of momenta through
track curvature rely on the DCH {\fmvrc and SVT} measurements in the 1.5-T magnetic
field generated by a superconducting solenoid.  The DCH {\fmvrc and SVT} measurements of
$dE/dx$ energy loss also contribute to charged-particle identification.

Surrounding the drift chamber is a novel detector of internally
reflected Cerenkov radiation (DIRC), giving charged-particle
identification in the central region of the detector. Outside the DIRC
is a highly segmented electromagnetic calorimeter (EMC) composed of
CsI(Tl) crystals. 
{\fmvrc The EMC is used to detect photons and neutral hadrons through shower shapes and is also used
to identify electrons.}
Finally, {\fmvsix the flux return of the superconducting coil surrounding the EMC 
is instrumented with resistive plate chambers interspersed with iron for the identification
of muons and neutral hadrons (IFR).}

A detailed Monte Carlo program based on
the GEANT4~\cite{ref:geant4} software package is used to simulate the
\babar\ detector response and performance.

%% file: sample.tex
From a sample of about 88 million $\FourS \to \B\Bb$ decays,
we select events in which one of the \B\ mesons is completely
reconstructed 
{\fmvcw in either a neutral or a charged hadronic final state,}
using the same criteria used for the
\babar\ $\sintwob$ measurement~\cite{ref:sin2b-babar} and for measurements of
\dM\ using hadronic final {\fmvsix states~\cite{ref:dM-babar-had}.
{\fmvcw Neutral-\B\ mesons are reconstructed in either a flavor (\bflav) 
or \CP\ (\bcp) eigenstate. The charged-\B-meson decays are
used as control samples in the cross-checks described in Sec.~\ref{sec:checks-data}.
The decay modes used for the flavor sample, the \CP\ sample, the control 
samples are displayed in Table~\ref{table:modes}.}
Details on
charged particle and neutral reconstruction, particle identification and reconstruction
of \B\ mesons can be found in Secs. II and III in Ref.~\cite{ref:babar-stwob-prd}}.

\begin{table}[h!t]
\centering
\caption{The flavor, \CP, and control {sample decay modes} used in this analysis. 
{\rncnov The $\jpsi$ is always identified in the $e^+e^-$ or $\mu^+\mu^-$ modes.
The $a_1^+$ is reconstructed only in $ \pip\pip\pim$.
The $\KS$ is identified in the $\pip\pim$ mode, except otherwise specified.}
{\ffsix All charge-conjugate decay modes are included implicitly.}
}
\label{table:modes}  
\vspace{0.1in}
\begin{tabular}{l|ll}\hline 
\\[-0.1in]
Samples & Decay modes & \\
\hline \hline \\[-0.1in]
$\bflav$ &$\Bz \to D^{*-}\pi^+(\rho^+,a_1^+)$&\\
&&$D^{*-}\to {\overline D}^0\pim$\\
&&$ {\overline D}^0\to \Kp\pim, \Kp\pim\piz,$ \\
& & $\qquad ~~~K^+\pim\pip\pim,$ \\
& & $\qquad ~~~\KS\pip\pim$\\
&$\Bz \to D^{-}\pi^+(\rho^+,a_1^+)$&\\
&&$D^{-}\to \Kp\pim\pim, \KS\pim$\\
&$\Bz \to \jpsi \Kstarz$&\\
&&$\Kstarz \to \Kp \pim$\\ \hline
\\[-0.1in]
$\bcp$&$\Bz\to \jpsi\KS$&\\
&&$\KS\to \pip\pim, \piz\piz$\\
&$\Bz\to \psitwos\KS$&\\
&&$\psitwos\to e^+e^-,\mu^+\mu^-,$\\
& & $\qquad ~~~~~~\jpsi\pip\pim$\\
&$\Bz\to \chic1\KS$&\\
&&$\chic1\to \jpsi\,\gamma$\\
&$\Bz\to \jpsi\KL$&\\ \hline
\\[-0.1in]
Control&$\Bu \to \overline{D}^{(*)0}\pip$&\\
& & $\overline{D}^{*0} \to \overline{D}^{0} \piz$ \\
&$\Bu \to \jpsi \Kp$&\\
&$\Bu \to \psitwos \Kp$&\\
&$\Bu \to \chic1 \Kp$&\\
&$\Bu \to \jpsi K^{*+}$&\\
&&$ K^{*+}\to \KS\pip$\\ \hline
\end{tabular}
\end{table}

We select \bflav\ and \bcp\ candidates by requiring that the
difference $\Delta E$ between their energy and the beam energy in the
center-of-mass frame be less than 
{\ffsix $3\sigma$ from zero, where $\sigma$ is the resolution on \de.}
The \de\ resolution ranges between 10 and 50 \mev\ depending on the decay mode.
For \bflav\ {modes} and \bcp\ modes involving \KS\ (\bcpks), the beam-energy
substituted mass must be greater than $5.2\gevcc$.  The beam-energy substituted mass is given by
\begin{equation}
\mes = \sqrt{\frac{( s/2 + \pvec_i\cdot \pvec_{\B})^2}{E_i^2} -
p_{\B}^2}~,
\end{equation}
where $s$ is the square of the center-of-mass energy,
$E_i$ and $\pvec_i$ are the total energy and the three-momentum of the
initial state in the laboratory frame, and $\pvec_{\B}$ is the
three-momentum of the \B\ candidate in the same frame.  In the case of
decays to $\jpsi\KL$ (\bcpkl), the \KL\ direction is measured but its momentum
is only inferred by constraining the mass of the $\jpsi\KL$ candidate
to the known \Bz\ mass.
 %Only the direction of \KL\ can be measured, so for the
As a consequence, there is only one
parameter left to define the signal region, which is taken to be $|\Delta E| < 10\mev$. 

\begin{figure}[!ht]
\centering
\epsfig{file=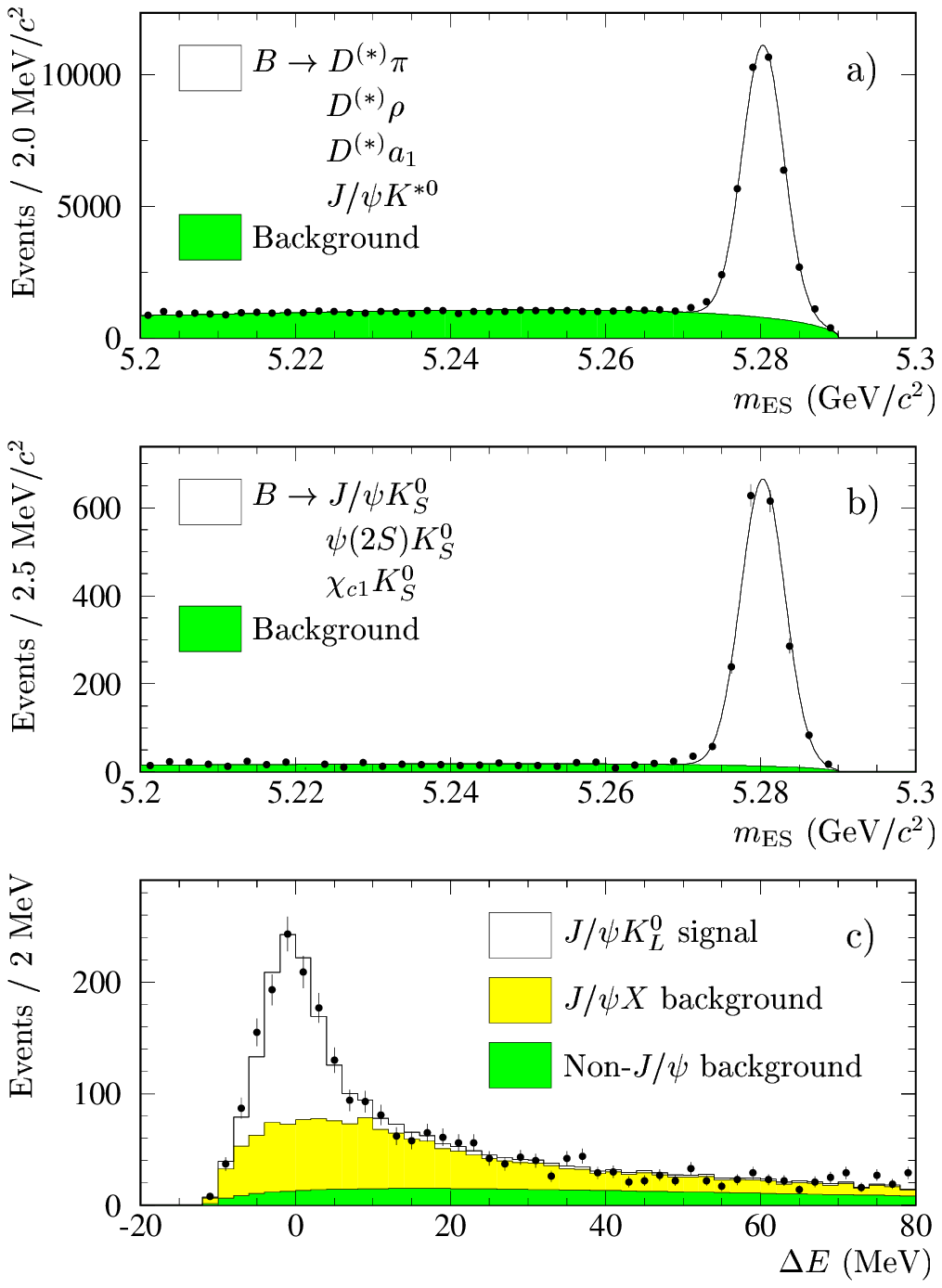,width=1.0\linewidth} \\
   \caption{Distributions for \bflav\ and \bcp\ candidates before vertex requirements: a) \mes\ for
\bflav\ states; b) \mes\ for $\Bz \to \jpsi \KS, \psitwos \KS, \chic1 \KS$ final states; and c)
$\Delta E$ for the final state $\Bz \to \jpsi \KL$.
In (a) and (b), the backgrounds are dominantly combinatorial.  In (c) there
are backgrounds from events containing a true $J/\psi$ but with a spurious
\KL.  Other background comes from events in which no true $J/\psi$ is 
present. 
\label{fig:sample}}
\end{figure}

Fig.~\ref{fig:sample} shows the \mes\ distribution for the \bflav\ and \bcpks\ samples and the $\Delta E$
distribution for the \bcpkl\ candidates, before the vertex
requirements (see Sec.~\ref{sec:decaytime}).
The combinatorial background in the \mes\ distributions 
 is described by the empirical ARGUS phase-space model~\cite{ref:argus} 
and the signal 
{\fmvcw by} a Gaussian distribution. The combinatorial background consists of 
{\fmvcw random combinations of tracks from continuum}
and \BB\ sources.
The former events are dominantly ``prompt'',
that is, the observed particles point back to the interaction point, whereas
the latter events are dominantly ``non-prompt'', with {\fmvsix particles pointing back
to separated} vertices.  Charmed particles, either from continuum or
from \B-meson decays, contribute to non-prompt background.
A small 
background due to other 
 \B\ decays (not shown in Fig.~\ref{fig:sample}) also peaks at the \B\ mass. The background in 
 the $\jpsi\KL$ channel receives contributions from other \B\ decays 
 with real \jpsi\ mesons in the final state, and {\un from events with fake
\jpsi\ mesons constructed from unassociated leptons or from misidentified
particles.}

After completely reconstructing one \B\ meson, the rest of the event
is analyzed to identify the flavor of the opposite \B\ meson and to
reconstruct its decay point, as described in Secs. \ref{sec:tagging}
and \ref{sec:decaytime}.

Using exactly the same requirements, {\fmvsix we analyze GEANT4-simulated samples}
 to check for any biases 
 in the {\fmvsix event selection and} extracted parameters.
The Monte Carlo samples are also used %to
in studies of detector response and to estimate some
background sources. The values of the {\fmvrc \B\ oscillation and \CP-, \T-, and \CPT- violating}
parameters assumed in the simulations are similar to those measured in the data. We use additional
samples with significantly different values to
check the reliability of the analysis in other regions of the
parameter space.

%% file: tagging.tex
The tracks that are not part of the fully reconstructed \B\ meson are
used to determine whether the \btag\ was a \Bz\ or \Bzb\ when it
decayed.  This determination cannot be done perfectly.  If the
probability of an incorrect assignment is $w$, an asymmetry that
depends on the difference between \Bz\ and \Bzb\ tags will be reduced
by a factor $D=1-2w$, called the dilution.  A neural network combining
the outputs of algorithms that evaluate the characteristics of each
event is used to take into account the correlations between the
different sources of flavor information and to estimate {\fmvrc \Bz\ and \Bzb\ mistag
probabilities} for each event. Based on {\fmvrc these values} and the source of flavor information, 
the event {\fmvrc is tagged and} assigned
to one of five mutually exclusive tagging categories.  The dilution for each
category is determined from the data, {\fmvrc as described in Sec.~\ref{sec:method}}. 
Grouping tags {\fmvcw into categories}, each with a
relatively narrow range in mistag probability, increases the overall
power of the tagging {\fmvrc while simplifying the studies of systematic uncertainties.}

Events with an identified primary electron or muon and a 
{\fmvrc kaon with the same charge,}
if present, are assigned to the {\tt Lepton} category.
Events with both an identified kaon and a 
{\fmvrc low-momentum (soft) pion
candidates} with opposite charge and similar
flight direction are assigned to the {\tt KaonI} category.
Soft pion {candidates} from \Dstarp\ decays are selected on the basis of
their momentum and direction with respect to the thrust axis of
\btag. Events with only an identified kaon are assigned to the {\tt KaonI} or 
{\tt KaonII} category depending on the estimated mistag probability.
Events with only a soft-pion candidate are assigned to the {\tt KaonII} 
category as well. 
The remaining events are assigned to {either} the {\tt Inclusive} or
{the} {\tt UnTagged} category based on the estimated mistag
probability. The {\tt UnTagged} tagging category has a mistag rate
set to 50\%, and therefore does not provide tagging information. 
It does, however, {\fmvrc increase} the sensitivity to the decay-rate difference
$\Delta\Gamma$ and allows the determination {\fmvrc from} the data of the
detector charge asymmetries, as described in Sec.~\ref{sec:method}.
This tagging algorithm is 
identical to that used in Ref.~\cite{ref:sin2b-babar}.

{\fmvrc We consider separate mistag probabilities for \Bz\ and \Bzb\ tags,
$\wa_{\tagv}$ and $\wa_{\tagbarv}$, in each tagging category $\alpha$. From these,
we define the average mistag probability
\mbox{$\wa = (\wa_{\tagv} + \wa_{\tagbarv})/2$} and the asymmetry in the mistag rates 
\mbox{$\dwa = \wa_{\tagv} - \wa_{\tagbarv}$}.}
A correlation between the {\fmvrc average} mistag rate and the \dt\
uncertainty $\sdt$ estimated event-by-event  (discussed in
Sec.~\ref{sec:decaytime}) is observed 
for kaon-based tags~\cite{ref:dM-babar-dstlnu,ref:babar-stwob-prd}. 
For a \deltat\ uncertainty less than \mbox{1.4\ps,} this correlation is found to be approximately
linear:

\begin{equation}
  \wa = \wa_0 + \wa_{\rm slope} \sdt~.
\label{eq:tagvtxcorr}
\end{equation}
All signal mistag parameters, $\wa_0$, $\wa_{\rm slope}$, and \dwa, are free in the global fit 
(11 in total since $w^{\tt Lepton}_{\rm slope}$ is assumed to be zero), and their results can 
be found in Table \ref{tab:mistag} in Sec.~\ref{sec:results}.

%% file: decaytime.tex
The time interval $\dt=t_{\rm rec} - t_{\rm tag}$ between the 
two \B\ decays is calculated from the measured separation \deltaz\ between the
decay {vertices of the reconstructed \brec\ meson and the \btag\ meson} 
along the $z$-axis, using the known boost of the \FourS\
resonance in the {\fmvcw laboratory, $\beta\gamma \approx 0.55$}, {the beam-spot size,
and the momentum of the fully reconstructed \B\ meson.}  The method is
the same as described in Sec.~V in Ref.~\cite{ref:babar-stwob-prd}.

An estimated error $\sigma_{\dt}$ on \dt\ is calculated for each
event. This error accounts for uncertainties in the track parameters
from the SVT and DCH hit resolution and {from} multiple scattering,
for the beam-spot size, and for effects from the
\B-flight length 
{\fmvrc transverse to the beam axis.}
However, it does not account for errors due to mistakes
of the pattern recognition system, wrong associations of tracks to
vertices, misalignment within and between the tracking devices,
inaccuracies in the modeling of the amount of material in the tracking
detectors, limitations in our knowledge of the {\fmvrc beam-spot position}, or uncertainty in the absolute $z$ scale. 
Most of the effects {\fmvcw that are} not explicitly accounted for in $\sdt$ are
absorbed in the \dt\ resolution function, described below.
Remaining systematic uncertainties are discussed in detail in Sec.~\ref{sec:systematics}.

We use only those events in which the vertices of the \brec\ and
\btag\ are successfully reconstructed and for which $| \dt | <
20$\ps and $\sdt<$\mbox{1.4\ps.}  The fraction of events in data
satisfying these requirements is about 85\%.  From Monte Carlo
simulation we find that the reconstruction efficiency does not depend
on the true value of \dt. The \rms\ $\Delta z$ resolution for 99.7\% of the
events used is about 160\mum\ \mbox{(1.0\ps)}, and is dominated by the resolution of the \btag\ vertex.

To model the \dt\ resolution we use the sum of three Gaussian
distributions (called core, tail and outlier
components) with different means and widths:
\bea
\resol(\delta t,\sdt
) & = &
f_{\rm core}  h_{\rm G}(\delta t;\delta_{\rm core}\sdt,S_{\rm core}\sdt) +  \nn \\
  & & f_{\rm tail} h_{\rm G}(\delta t;\delta_{\rm tail}\sdt,S_{\rm tail}\sdt) + \nn \\
  & & f_{\rm out} h_{\rm G}(\delta t;\delta_{\rm out},\sigma_{\rm out}) 
\label{eq:resol}
\eea
where
\bea
  h_{\rm G}(\delta t;\delta,\sigma) & = & 
       \frac{1}{\sqrt{2\pi}\sigma} e^{ -(\delta t-\delta)^2/(2\sigma^2)  }~.
\label{eq:resolGauss}
\eea
Here
\mbox{$\delta t = \dt-\dt_{\rm true}$} represents the reconstruction
error and $f_{\rm core}=1-f_{\rm tail}-f_{\rm out}$.  
We incorporate the last Gaussian {\fmvrc distribution} in
Eq.~(\ref{eq:resol}) without reference to $\sdt$ since the outlier
component is not expected to be well described by the estimated uncertainty.
{The widths of the first two Gaussian components are given by $\sigma_{\Delta t}$ multiplied by 
two independent scale factors,
$S_{\rm core}$ and $S_{\rm tail}$,} to accommodate an overall
underestimate ($S>1$) or overestimate ($S<1$) of the 
errors. The core and tail Gaussian distributions are allowed to have
non-zero means ($\delta_{\rm core}\sdt$ and $\delta_{\rm tail}\sdt$) to account for residual biases due to daughters of
long-lived charm particles included in the $B_{\rm tag}$ vertex.
Separate means are used for the core distribution of each tagging
category.  These means are scaled by $\sdt$ to account for a
correlation %observed in Monte Carlo simulation 
between the mean of the
$\delta t$ distribution and $\sdt$~\cite{ref:dM-babar-dstlnu,ref:babar-stwob-prd}.  This correlation is
found to be approximately linear for 
$\sigma_{\dt}$ less than
\mbox{1.4\ps.}
The non-zero means of the resolution function introduce an asymmetry
into the otherwise symmetric \dt\ distributions.  {All other parameters 
of the resolution function {are} taken to be independent of the tagging category.}
We find that the three parameters
describing the outlier Gaussian component are {strongly} correlated among
themselves and with other resolution function parameters. Therefore, we fix
the outlier bias $\delta_{\rm out}$ and width $\sigma_{\rm out}$ to 0\ps and 8\ps, 
respectively, and vary them through a wide range to evaluate systematic
uncertainties. 
The outlier {\fmvrc Gaussian distribution
accounts} for less than 0.3\% of the reconstructed vertices.

In simulated events, we
find no significant differences between the
\dtresolutionfunction~of the \bflav, \bcpks, and \bcpkl\ samples.
This is expected since the $B_{\rm tag}$ vertex
precision dominates the \deltat\ resolution. Hence, the same resolution
function is used for all modes. Possible residual differences are taken into account
in the evaluation of systematic errors described in Sec.~\ref{sec:systematics}.

The resulting signal resolution function is
described by a total of 12 parameters, 
$S_{\rm core}$,
$\delta_{\rm core}^{\tt Lepton}$, 
$\delta_{\rm core}^{\tt Kaon I}$,
$\delta_{\rm core}^{\tt Kaon II}$, 
$\delta_{\rm core}^{\tt Inclusive}$,
$\delta_{\rm core}^{\tt UnTagged}$,
$f_{\rm tail}$, 
$\delta_{\rm tail}$,
$S_{\rm tail}$, 
$f_{\rm out}$, 
$\delta_{\rm out}$, 
$\sigma_{\rm out}$,
ten of which are {\ffsix free}  in the final fit.  

As a cross-check, we use an alternative resolution function that is
the sum of a single Gaussian distribution (centered at zero), the
same Gaussian convolved with a one-sided exponential to describe the
core and tail parts of the resolution function, and a single Gaussian
distribution to describe the outlier component~\cite{ref:dM-babar-dstlnu}. The exponential
component is used to accommodate the bias due to tracks from charm decays 
{\fmvrc originating from the \btag}.
The exponential constant is scaled by
$\sdt$ to account for the previously described correlation between the
mean of the $\delta t$ distribution and $\sdt$. In this case, each
tagging category has a different core component fraction and
exponential constant.

%% file: method.tex
We perform a single, unbinned maximum-likelihood fit to all \bflav,
\bcpks, and \bcpkl\ samples. Each event is characterized by the {\fmvrc following quantities:}
\begin{enumerate}[topsep=2pt, partopsep=2pt, parsep=1pt, itemsep=1pt, label=\emph{\roman*})]
\item assigned tag category $\alpha$ $\in$ \{{\tt Lepton, KaonI, KaonII, Inclusive, UnTagged}\}; 
\item tag-flavor type \tagip~$\in$ \{\tagv, \tagbarv\}, i.e., the tagging state is ostensibly a \Bz or \Bzb,
unless it is untagged; 
\item reconstructed event type \recip~$\in$ \{\flavv, \flavbarv, \cpks, \cpkl\}, 
i.e., the reconstructed state is ostensibly a \Bz, \Bzb, 
or a \CP\ eigenstate. \fmvrf{Treating \KS\ and \KL\ as if they were \CP\ eigenstates introduces
effects that are negligible on the scale of the statistical and systematic uncertainties of this analysis;}
\item the decay-time measurement \dt\ and its estimated error $\sdt$;
\item a variable $\zeta$ used to assign the probability that the event is signal or background.
Either $\zeta$ is \mes\ (for flavor eigenstates and \CP\ eigenstates 
with \KS ) or it is \de\ (for \CP\ eigenstates with \KL).
\end{enumerate}
The likelihood function is built from time distributions that depend on
whether the event is signal or any of a variety of backgrounds
(together specified by the \mbox{index $j$}), on the tag category, on the
tag flavor, and on the type of reconstructed final state.}
The contribution of a single event to the log-likelihood is
\begin{equation}
\log \left[ \sum_j {\cal F}^{\alpha,j}_{\reci}(\zeta) {\cal H}_{{\tagi,\reci}}^{\alpha,j}(\dt,\sdt) \right]~.
\label{eq:lik}
\end{equation}
For a given reconstructed event type \recip\ and tagging category
$\alpha$, $ {\cal F}^{\alpha,j}_{{\reci}}(\zeta)$ gives the {\fmvsix probability}
that the event belongs to the signal or any of the various backgrounds
denoted by $j$. Each such {component} has its own {\fmvrc probability density function} (PDF) $ {\cal
H}_{{\tagi,\reci}}^{\alpha,j}(\dt,\sdt)$, which depends as well on the
particular tag flavor \tagip.  This distribution is the convolution of a
tagging-category-dependent time distribution 
${ H}^{\alpha,j}_{{\tagi,\reci}}(\dt_{\rm true})$  
with {\ffsix a \dtresolutionfunction\ $\resol^{\alpha,j}(\delta t,\sdt)$ of the form 
given in Eq.~(\ref{eq:resol}), but with parameters that depend on the 
tagging category $\alpha$ and on the signal/background nature of the \mbox{event $j$:}}

\begin{widetext}
\bea
  {\cal H}^{\alpha,j}_{{\tagi,\reci}}(\dt,\sdt) & = & \int_{-\infty}^{+\infty} {\rm d}(\dt_{\rm true}) 
  \resol^{\alpha,j}(\dt-\dt_{\rm true},\sigma_{\dt}
) { H}^{\alpha,j}_{{\tagi,\reci}}(\dt_{\rm true}) ~,
\label{eq:intensitiesAnaResol}
\eea
where
\begin{eqnarray}
  { H}^{\alpha,j}_{{\tagi,\reci}}(\dt_{\rm true}) & = & \rho_{{\reci}}^j \left\{ 
       \tau^{\alpha,j}_{{\tagi}} (1 - \waj_{{\tagi}}) { h}^{ j}_{{\tagi,\reci}}(\dt_{\rm true}) + 
       \tau^{\alpha,j}_{{\overline{{\tagi}}}} \waj_{{\overline{{\tagi}}}} 
        { h}^{ j}_{{\overline{{\tagi}}},\reci}(\dt_{\rm true})
        \right\}~.~~~
\label{eq:intenmistag}
\end{eqnarray}
\end{widetext}
{\ffsix 
Here ${ h}^j_{{\tagi,\reci}}(t)$ represents the time
dependence ${\rm d}N / {\rm d}t$  given in 
Eqs.~(\ref{eq:pref2}) - (\ref{eq:s}), with
$t \equiv \dt_{\rm true}$.
}
{\rncnov
We indicate by
$\waj_{{\tagi}/{\tagbari}}$ the mistag fractions for category
$\alpha$ and component $j$. The index \tagbarip\ denotes the
opposite flavor to that given by \tagip. For events falling into
tagging category {\tt UnTagged} {\fmvsix we define $\waj_{{\tagi}/{\tagbari}}$ to be 1/2.}
The efficiency $\tau^{\alpha,j}_{\tagi}$ is the 
probability that an event whose signal/background
nature is $j$ and whose true tag flavor is \tagip\ will be assigned to
category $\alpha$, regardless of whether the flavor assigned is
correct or not.  The efficiency $\rho^j_{{\reci}}$ is the probability
that an event whose signal/background nature is indicated by $j$ and
whose true {\fmvsix reconstructed} character is \recip\ will, in fact, be reconstructed.
}
{\fmvsix For non-\BB\ background sources, where the meaning of true \tagip\ and \recip\ is ambiguous, this
provides an empirical description of the efficiencies as well as the mistag fractions.}
}

\subsection{PDF Normalization}

Every reconstructed event, whether signal or background occurs at some
time $\Delta t_{\rm true}$, so

\bea
\int_{-\infty}^{+\infty}\, {\rm d}(\Delta t_{\rm true})
    {h}^{j}_{{\tagi,\reci}}(\dt_{\rm true}) & = & 1 ~,~~~
\label{eq:norm}
\eea
for each value of \recip, \tagip~and $j$.
Moreover, every event is assigned to some tagging category (possibly {\tt UnTagged}); thus
\begin{equation}
\sum_\alpha \tau^{\alpha,j}_{{\tagi}}=1
\end{equation}
for each value of \tagip\ and $j$.
It follows then that the normalization of  $ {H}^{\alpha,j}_{{\tagi,\reci}}(\dt_{\rm true})$ is
\begin{equation}
\sum_\alpha\sum_{{\tagi}} \int_{-\infty}^{+\infty}\, {\rm d}(\Delta t_{\rm true})
    {H}^{\alpha,j}_{{\tagi,\reci}}(\dt_{\rm true})=2 \rho^j_{{\reci}}~.
\label{eq:norm2}
\end{equation}
In this analysis the nominal normalization of ${\cal H}^{\alpha,j}_{{\tagi,\reci}}(\dt,\sdt)$ is the same
as $ {H}^{\alpha,j}_{{\tagi,\reci}}(\dt_{\rm true})$,
but fits with normalization in the interval $[-20,20]$\ps have been also performed as a cross-check to evaluate 
possible systematic effects.

\subsection{Signal and Background Characterization}

The function ${\cal F}^{\alpha,j}_{{\reci}}(\zeta)$ in Eq. (\ref{eq:lik})
describes the signal or background probability of observing a
particular value of $\zeta$. It satisfies
\begin{equation}
\int_{\zeta_{\rm min}}^{\zeta_{\rm max}} {\rm d}\zeta \sum_j {\cal F}^{\alpha,j}_{{\reci}} (\zeta)  =  1~,
\end{equation}
where $[\zeta_{\rm min},\zeta_{\rm max}]$ is the range of \mes\ or \de\ values
used for analysis.  

For \bflav\ and \bcpks\ events, the \mes\ shape is described with a
single Gaussian distribution for the signal and an ARGUS
parameterization for the background~\cite{ref:babar-stwob-prd}. 
Based on these fits, an event-by-event
signal probability \palphar\
can be calculated for
each tagging category $\alpha$ and sample \recip. {Since} we do not expect
signal probability differences between
\Bz\ and \Bzb, the \mes\ fits are performed to \flavv\ and \flavbarv\ events together. 
{\fmvcw The fits to $\Bz \to \psitwos\KS$ and $\Bz \to \chicone\KS$  are performed
without subdividing by tagging category,
due to the lack of statistics and the high purity of the samples.}
{\fmvrc We distinguish three different background components: peaking background events, which have
the same \mes\ behavior as the signal; a zero-lifetime (prompt) combinatorial component; and a non-zero-lifetime
(non-prompt) combinatorial background.} 
The component fractions ${\cal F}^{\alpha,j}_{{\reci}}(\mes)$ are {\fmvrc then}
($j={\rm sig, peak,} k$)
\begin{eqnarray}
{\cal F}^{\alpha,\rm sig}_{{\reci}}(\mes) & = &
\left[ 1-f^{\alpha,\rm peak}_{{\reci}} \right] \palphar, \nn \\
{\cal F}^{\alpha,\rm peak}_{{\reci}}(\mes) & = &
   f^{\alpha,\rm peak}_{{\reci}} \palphar, \nn \\
{\cal F}^{\alpha,k}_{{\reci}}(\mes) & = &
\left[ 1 - \palphar \right] f^{\alpha,k}_{{\reci}},
\end{eqnarray}
where $k$ indexes the various combinatorial {\fmvsix ($k$ = prompt, non-prompt)} background components, {\un and}
\begin{equation}
\sum_{k}  f^{\alpha,k}_{{\reci}} =1~.
\end{equation}
The fraction $f^{\alpha,\rm peak}_{{\reci}}$ of the signal Gaussian
distribution is due to backgrounds that peak in the same {\un regions} as the signal, and is 
determined from Monte Carlo simulation~\cite{ref:babar-stwob-prd}. The estimated
contributions are $(1.5\pm0.6)\%$, $(0.28\pm0.11)\%$, $(1.8\pm0.6)\%$,
$(1{+3\atop -1})\%$, and $(3.5\pm1.4)\%$ for the \bflav, $\jpsi\KS\ (\KS \to
\pipi)$, $\jpsi\KS\ (\KS \to \ppz)$, $\psitwos\KS$, and $\chicone\KS$
channels, respectively. A common peaking background fraction is
assumed for all tagging categories within each decay mode.  
We {\un also} assume a common prompt fraction for all tagging categories for each \bcpks\
decay channel. 
{\un Since the
\bflav\ sample is large  and there are significant differences in the background
levels for each tagging category, $f^{\alpha,\rm prompt}_{\flavv}=f^{\alpha,\rm prompt}_{\flavbarv}$ is
allowed to depend on the tagging category.}  Note that the parameters
of the ${\cal F}^{\alpha,\rm sig}_{{\reci}}(\mes)$ functions, determined from
a set of separate unbinned maximum-likelihood fits to the \mes\
distributions, are  fixed in the global fit.

For \bcpkl\ events the background level is
higher than it is for \bcpks, with significant non-combinatorial components~\cite{ref:babar-stwob-prd}.  A binned
likelihood fit to the \de\ spectrum in the data is used to determine
the relative amounts of signal and background from $\B \to \jpsi X$
(e.g., $\jpsi K^*$)
events and from events {with} {misreconstructed $\jpsi \to \ell^+ \ell^-$
candidates} (non-\jpsi\ background). In these fits, the signal and $\B
\to \jpsi X$ background distributions are obtained from inclusive-\jpsi\ Monte Carlo {samples}, 
while the non-\jpsi\ distribution is obtained from
the \jpsi\ dilepton-mass sideband.  The Monte Carlo simulation is also
used to evaluate the channels that contribute to the $\B \to \jpsi X$
background. 
The fit is performed separately for \KL\ candidates reconstructed in the EMC
and in the IFR, and for \jpsi\ candidates reconstructed in the \epem\ and \mumu modes,
since there are differences in purity and background composition.
{\fmvcw Candidates reconstructed in both IFR and EMC are considered as belonging to the IFR category
because of its better signal purity}.
The different inclusive-\jpsi\ backgrounds from Monte Carlo are then 
normalized to the \jpsi\ background fraction extracted from the \de\ fit in the data.
The normalization to the data is performed separately for lepton-tagged and non-lepton-tagged events to account
for the observed differences in flavor-tagging efficiencies between 
the \jpsi\ sideband events and the \bflav\ and
inclusive-\jpsi\ Monte Carlo events.
In addition, some of the decay modes in the inclusive-\jpsi\ background have \CP\ content. 
\remark{ Deleted formula }
The same PDFs are used to describe {\fmvrc the \de\ shape} for \jpsi\ candidates in 
the $\mu^+\mu^-$ and $e^+e^-$ channels.  However, different  PDFs are
used for \KL\!\!s observed in the IFR and in the EMC. Separate \de\ PDFs are used for
$\jpsi\KL$ (signal), $\jpsi\KS$ background, $\jpsi X$ background
(excluding $\jpsi\KS$), and non-\jpsi background.

\subsection{Efficiency Asymmetries}

For each signal or background $j$, the average reconstruction efficiencies
$\rho^j = (\rho^j_{\flavv}+\rho^j_{\flavbarv})/2$, $\rho^j_{\cpks}$, and $\rho^j_{\cpkl}$ are 
absorbed {\fmvcw into} the fractions of reconstructed events falling into the different
signal and background classes.  In contrast, because all events fall into some tagging
category (including {\tt UnTagged}), the {\un average 
tagging efficiencies} $\tau^{\alpha,j} = (\tau^{\alpha,j}_{\tagv}+
\tau^{\alpha,j}_{\tagbarv})/2$ are meaningful, 
and {\rncnov the fraction of untagged signal events plays an important role.}
The asymmetries in the efficiencies, 
\begin{eqnarray}
\nu^j&=&\frac{\rho^j_{\flavv}-\rho^j_{\flavbarv}}{\rho^j_{\flavv}+\rho^j_{\flavbarv}}~,\nonumber\\
\mu^{\alpha,j}&=&\frac{\tau^{\alpha,j}_{\tagv}-\tau^{\alpha,j}_{\tagbarv}}{\tau^{\alpha,j}_{\tagv}+
    \tau^{\alpha,j}_{\tagbarv}}~,
\label{eq:numudef}
\end{eqnarray}
need to be determined precisely, because they might otherwise mimic fundamental
asymmetries we seek to measure. In Appendix \ref{appendix:asymmetries}
we illustrate how the use of the untagged sample makes it possible to determine
the asymmetries in the efficiencies. Note that asymmetries due to differences
in the magnitudes of the decay amplitudes, $|A_{\flavv}| \ne |\overline{A}_{\flavbarv}|$ 
and $|A_{\tagv}| \ne |\overline{A}_{\tagbarv}|$, 
cannot be distinguished {\fmvcw from asymmetries} in the efficiencies, 
and thus are absorbed in the $\nu$ and $\mu$ parameters.

We determine the average tagging efficiencies $\tau^{\alpha,j}$ by counting the number 
of events falling {\fmvcw into} different tagging categories,
without distinguishing where an event is signal or {\fmvcw background 
(i.e., $\tau^{\alpha,j} \equiv \tau^{\alpha}$), since} for each tagging
category $\alpha$ the $j$ component
dependence is absorbed {\fmvcw into} the fractions of events falling into the different signal and background components.
For signal events, the parameters $\nu^{\rm sig}$
and $\mu^{\alpha,\rm sig}$ are included as free parameters in
the global fit, and are assumed to be the same for all \Bz\ peaking
background sources.
For \Bu\ peaking
background components, $\nu^{\rm peak}$ and $\mu^{\alpha,\rm peak}$ are fixed to the values extracted from a
previous unbinned maximum-likelihood fit to the tagged and untagged \dt\
distributions of \Bu\ data used as control samples, described in
Sec.~\ref{sec:sample}. For combinatorial background sources 
the {\fmvrc $\nu$ and $\mu$ parameters} are neglected.

\subsection{Mistags and \dtResolutionFunction}

For signal events, a common set of mistag and \dtresolutionfunction\ parameters, 
{independent of the particular 
fully reconstructed state}, is assumed. This
assumption is supported {\fmvcw by Monte} Carlo studies.  

Peaking backgrounds originating 
from  \Bz\ decays are assumed to have the same resolution function 
and mistag parameters as the signal. For \Bu\ peaking backgrounds 
{\fmvrc we assume} the same resolution function as for signal, but the mistag
parameters are fixed to the values extracted from the 
{\fmvrc same maximum-likelihood fit to the \Bu\ data used to extract the parameters
$\nu^{\rm peak}$ and $\mu^{\alpha,\rm peak}$, as described above}.

For combinatorial background components 
(prompt and non-prompt components in the \bflav\ and \bcpks\ samples
and the non-\jpsi\ background in the \bcpkl\ sample) we use
an empirical description of the mistag probabilities and \dt\ resolution, allowing
various intrinsic time dependences. 
The parameters $\dwa$ and $\wa_{\rm slope}$ are fixed to zero, and
the resolution model uses core and outlier Gaussian distributions.
The fractions of prompt and non-prompt components and the {\fmvrc lifetime}
of the non-prompt component in the non-\jpsi\ background are
fixed to the values obtained from an external fit to the time
distribution of the \jpsi\ dilepton-mass sideband.  
\subsection{Free Parameters for the Nominal Fit}

The {\fmvrc aim} of the fit is to obtain simultaneously
\sgndGoverG, \absqop, \reZparflat, and \imZ, assuming $\Rcp=1$. 
The parameters \imlambcpflat\ and \dM\ are also free in the fit to account for
possible correlations and to provide an additional cross-check of the
measurements.
The average \Bz\ lifetime $\taub\equiv 1/\G$
is fixed to the PDG value, 1.542\ps~\cite{ref:pdg2002}.  
As a cross-check we also perform fits allowing
$\Rcp$ and $\G$ to vary.  All these physics parameters are, by
construction, common to all samples and tagging categories, although
the statistical power for determining each parameter comes from a
particular combination of samples or \dt\ dependences, as discussed in
Sec.~\ref{sec:decayrates}.

{\fmvsix The terms proportional to the real parts of the {\fmvrc \dckm-decay} parameters
are small since \relambflavbare\ and \relambbarflavbarbare\ occur only
multiplied by other small parameters (see Tables~\ref{table:cplus}-\ref{table:s}), 
and are therefore neglected in the nominal fit model. 
Fixing $|\overline{A}_{\flavv}/A_{\flavv}|=0.02$, our best estimate from $|V_{ub}^*V_{dc}^{}/V^*_{cb}V_{ud}^{}|$, 
we fit for the parameter $ \imlambflavbare /|\lambflavbare|$, and vary
separately $ \imlambbarflavbarbare /|\lambbarflavbarbare|$, keeping
$|\lambbarflavbarbare| = |\lambflavbare| |p/q|^2$. We do not require
$| \imlambflavbare /\lambflavbare|\leq 1$. {\fmvcw Thus,} there are two free parameters
{\fmvrc associated to \dckm\ decays}, plus one fixed magnitude.
}

We treat $\lambtagbare$ and $\lambbartagbarbare$ similarly.
{\fmvcw Since} there is no interference betweeen \Bz\ and \Bzb\ semileptonic
decays, we set $\lambtagbare=0$, $\lambbartagbarbare=0$ for the ${\tt
Lepton}$ tagging category.  For the other tagging categories we assume
common values of the {\fmvrc \dckm-decay} parameters.  We assign a systematic error by
varying 
$|\overline{A}_{\flavv}/A_{\flavv}|$ and $|\overline{A}_{\tagv}/A_{\tagv}|$
by 100\% and scanning all possible combinations of the phases
(Sec.~\ref{sec:systematics}).  With a larger data sample, direct
determination of the {\fmvrc \dckm-decay} parameters might be advantageous.  With the
current sample, absorbing some of the variation into the systematic uncertainty 
suffices {\fmvrc to prevent effects induced by \dckm\ decays being misinterpreted as symmetry violations}.

The total number of parameters that are free in the fit is 58, of
which 36 parameterize the signal: physics parameters (4), cross-check
physics parameters (2), single effective imaginary parts of the
\dckm-decay phases (4), resolution function (10), mistag
probabilities (11), and differences in the fraction of \Bz\ and \Bzb\ mesons 
that are tagged and reconstructed (5).
{\fmvcw The remaining 22 parameters are used to model the combinatorial backgrounds:
resolution function (3), mistag fractions (8), fractions of prompt components (9) and
the effective lifetime of the non-prompt contributions (2).}

The \dt\ distributions, the asymmetries, the physics parameters
\sgndGoverG, \absqop, \reZparflat, and \imZ\, and the
cross-check parameter \imlambcpflat\ were kept hidden until the analysis
was finished. However, the parameter \dM, the residual \dt\ distributions and asymmetries,
the statistical errors, and changes in the physics parameters due to
changes in the analysis were not hidden.

%% file: results.tex
We extract the parameters \sgndGoverG, \absqop, \reZparflat, \imZ, \mbox{\imlambcpflat},
\dM, the parameters for \dckm\ decays, 
the signal mistag probabilities, resolution-function and $\nu$ and $\mu^\alpha$ parameters, and the empirical 
background parameters with the likelihood function
described in Sec.~\ref{sec:method}. In Table \ref{tab:signalyields} we list the signal yields
in each tagging category after vertex requirements. The purities 
(estimated {\fmvcw from the \mes\ fits}
for non-\bcpkl\ samples 
and {\fmvcw in the region} $| \Delta E | < 10$ \mev\ for \bcpkl\ events), averaged over
tagging categories, are 82\%, 94\%, and 55\%, for \bflav, \bcpks, and
\bcpkl\ candidates, respectively. The fitted signal mistag probabilities and
resolution-function parameters are shown in Tables
\ref{tab:mistag} and \ref{tab:resolution}. The values of the
asymmetries in reconstruction and tagging efficiencies are summarized
in Table \ref{tab:munu}. There is good agreement with the
asymmetries extracted with the counting-based approach outlined in
Appendix \ref{appendix:asymmetries}.

\begin{table}[htb!]
\centering
\caption{Signal event yields {\fmvrc after vertex requirements}, obtained 
from the \mes\ fits for the \bflav\ and \bcpks\ samples.  
For the \bcpkl\ sample, the signal {\fmvrc yields are obtained using
the signal fractions determined from the fit to the \de\ distributions, and are
quoted for events satisfying} $| \Delta E |<10$ \mev.\label{tab:signalyields}}
\vspace{0.1in}
\setlength{\extrarowheight}{3pt}
\begin{tabular}{l|ccc|ccc|ccc} \hline
     &  \multicolumn{3}{c|}{\bflav} & \multicolumn{3}{c|}{\bcpks} & \multicolumn{3}{c}{\bcpkl} \\
\hline
 Tag &   \Bz & \Bzb & Tot & \Bz & \Bzb & Tot & \Bz & \Bzb & Tot \\
\hline \hline
 {\tt Lepton}      & 1478 & 1419 & 2897 & 96  & 98  & 194 & 35 & 35 & 70 \\
 {\tt Kaon I}    & 2665 & 2672 & 5337 & 154 & 175 & 329 & 74 & 65 & 139 \\
 {\tt Kaon II} & 3183 & 2976 & 6159 & 181 & 188 & 369 & 85 & 66 & 151 \\
 {\tt Inclusive}       & 3197 & 3014 & 6211 & 184 & 172 & 356 & 78 & 72 & 150 \\
 {\tt UnTagged}    & \multicolumn{3}{r|}{{ 10423}} & \multicolumn{3}{r|}{{ 585}} & \multicolumn{3}{r}{{ 260}} \\
\hline
\end{tabular}
\end{table}

\begin{table}[!htb] 
\centering
\caption{Average tagging efficiencies {\fmvrc after vertex requirements} and signal mistag parameters for each
  tagging category $\alpha$ as extracted from the 
  maximum-likelihood fit that allows for \CPT\ violation. Uncertainties are statistical only.}
\vspace{0.1in}
\label{tab:mistag} 
\input{mistag-table}
\end{table}

\begin{table}[!htb] 
\centering
\caption{Signal \dtresolutionfunction\ parameters as extracted
  from the maximum-likelihood fit that allows for \CPT\ violation. Uncertainties are
  statistical only.}
\vspace{0.1in}
\label{tab:resolution} 
\input{resolution-table}
\end{table}

\begin{table}[!htb] 
\centering
\caption{Values of the signal $\Bz\Bzb$ differences in reconstruction ($\nu^{\rm sig}$) and 
tagging ($\mu^{\alpha,\rm sig}$) efficiencies as extracted from the 
maximum-likelihood fit that allows for \CPT\ violation. 
The results are compared with those obtained with a counting-based method described in Appendix~\ref{appendix:asymmetries}.}
\vspace{0.1in}
\label{tab:munu} 
\input{munus-table}
\end{table}

The values of {\fmvcw the parameters} \sgndGoverG, \absqop, \reZparflat, and \imZ\ extracted
from the fits are given in Table \ref{tab:phys-param}.  
{\fmvsix The fitted \imlambcpflat, \dM, and}
effective {\fmvrc \dckm-decay} parameters are also
indicated. All these results can be compared to those obtained when
the fit is repeated assuming \CPT\ invariance. The change
in the effective {\fmvrc \dckm-decay} parameters between
the two fits is due to the large correlation of these parameters with
the \CPT-violating parameter \imZ. The fitted value of \dM\ agrees with
recent \B-factory measurements~\cite{ref:dM-babar-had,ref:dM-babar-dstlnu,ref:dM-babar-dilep,ref:dM-belle}, 
and remains unchanged between the two fits. The fit result for \imlambcpflat\ when we 
assume \CPT\ invariance 
agrees with our $\sintwob$ measurement
based on the same data set~\cite{ref:sin2b-babar}. When we allow for
\CPT\ violation, \imlambcpflat\ increases by $+0.011$, 
\fmvrf{equal to 15\% of the statistical uncertainty on \imlambcpflat, which is consistent with the
statistical correlations observed in the fit with \z\ free.}
The correlation coefficients among all physics and cross-check physics parameters
are shown in Table \ref{tab:correlation}. 
{\fmvsix The largest observed correlation (17\%) appears between \imZ\ and \imlambcpflat.}
Table \ref{tab:correlation-top} shows 
the largest statistical correlations of the physics parameters with any other 
free parameter in the fit. {\fmvrc Note that the variables \absqop\ and 
$\nu^{\rm sig}$ are significantly correlated, as are \imZ\ and the {\fmvrc \dckm-decay} parameters}.
\fmvrf{We do not evaluate the  full systematic errors for \dM\ and \imlambcpflat\ so
these measurements do not supersede previous \babar\ measurements for these
quantities.}

\begin{table}[!htb] 
\centering
\caption{Physics parameters extracted  from the maximum-likelihood fits both allowing for \CPT\ violation and excluding it.
The free {\fmvrc \dckm-decay} parameters are also indicated.
Errors are statistical only.} 
\label{tab:phys-param} 
\vspace{0.1in}
\input{physpar-table}
\end{table}

\begin{table}[!htb] 
\centering
\caption{Correlation (in \%) among all the physics parameters extracted from the 
simultaneous maximum-likelihood fit to the \bflav\ and \bcp\
samples. 
}
\label{tab:correlation}  
\vspace{0.1in}
\input{correlation-table}
\end{table} 

\begin{table}[!htb] 
\centering
\caption{The largest correlations of each physics parameter with other free parameters
of the  maximum-likelihood fit.}
\label{tab:correlation-top}  
\vspace{0.1in}
\input{correlation-table-top}
\end{table}

\begin{figure}[!ht]
\centering
\begin{tabular} {c}  
\epsfig{file=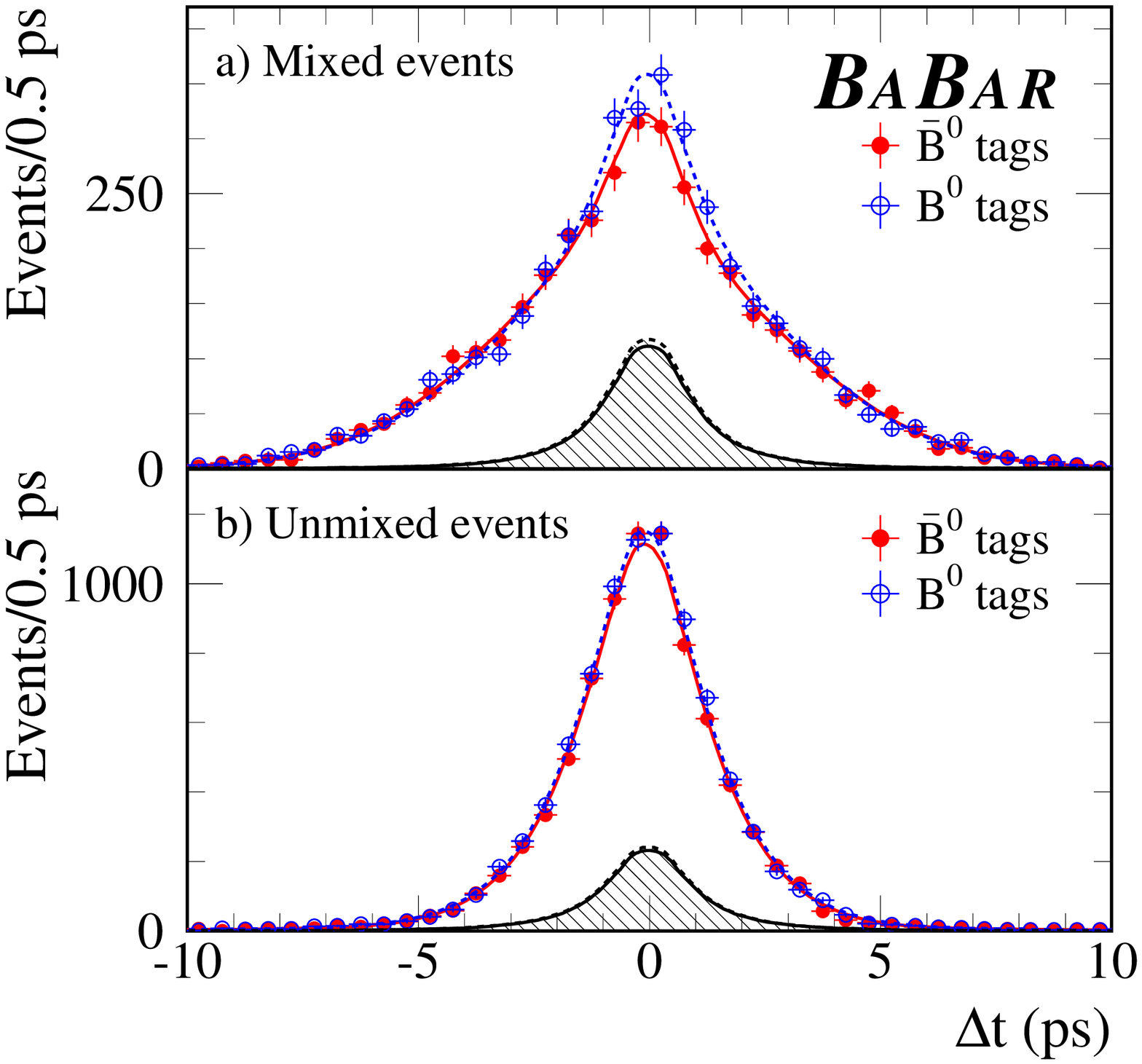,width=1.0\linewidth} 
\end{tabular} 
\caption{The \dt\ distributions for (a) mixed and (b) unmixed  \bflav\ events with a \Bz\ tag or with a \Bzb\ tag
in the signal region, $\mes>5.27$ \gevcc. The solid (dashed) curves represent
the fit projection in \dt\ based on the individual signal {\fmvrc and background} 
probabilities and {\fmvrc the} event-by-event \dt\ uncertainty for \Bzb(\Bz) tags. 
The shaded area shows the background contribution to the distributions. 
\label{fig:dtbflav}}
\end{figure}

\begin{figure}[!ht]
\centering
\begin{tabular} {c}  
\epsfig{file=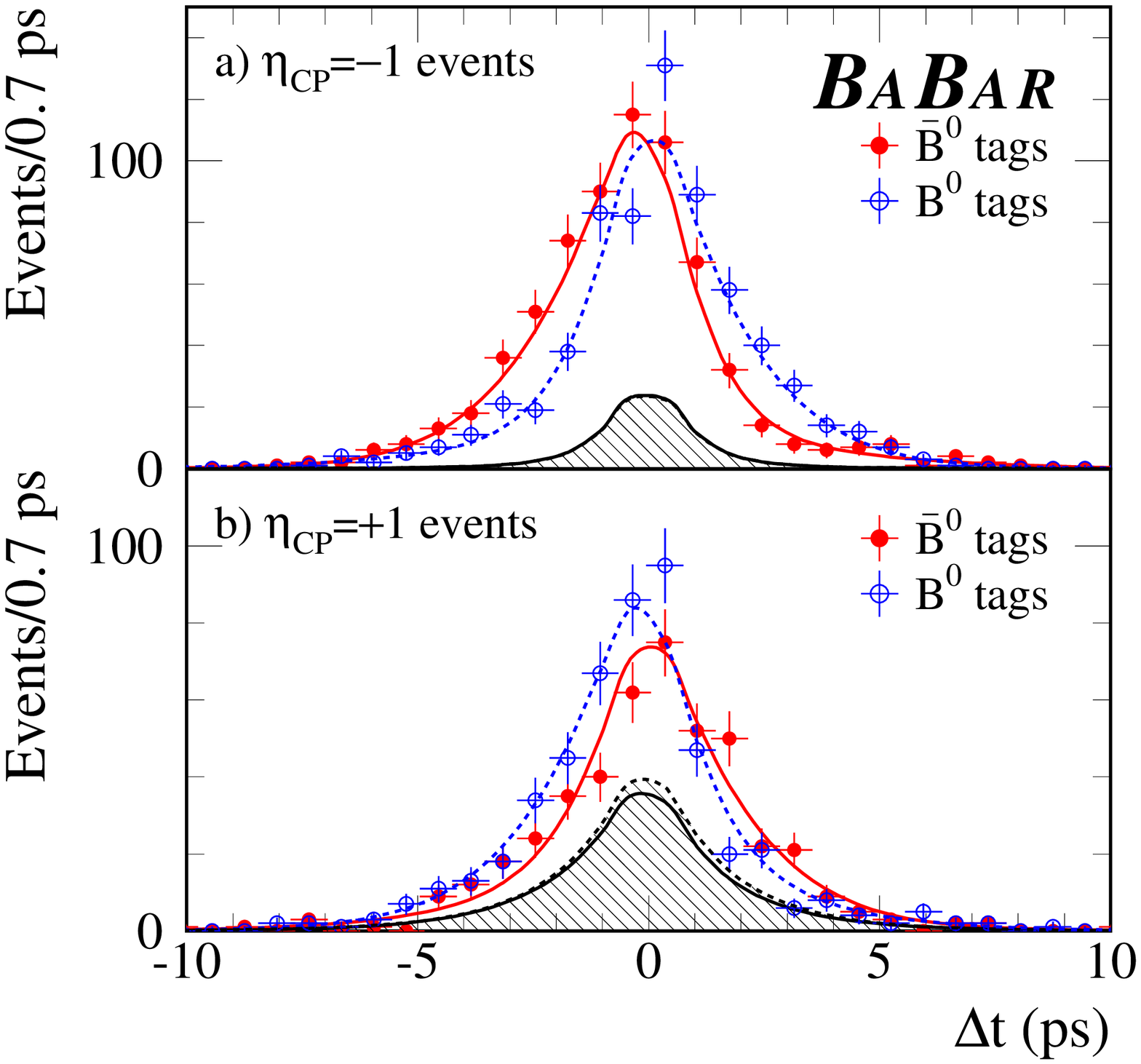,width=1.0\linewidth}
\end{tabular}   
\vspace{-0.5in}
\caption{\fmvrf{The \dt\ distributions} for (a) \bcpks\ and (b) \bcpkl\ events with a \Bz\ tag or with a \Bzb\ tag
in the signal region, $\mes>5.27$ \gevcc\ for \bcpks\ candidates 
and $| \Delta E | < 10$ \mev\ for \bcpkl\ events. The solid (dashed) curves represent
the fit projection in \dt\ based on the individual signal {\fmvrc and background} probabilities 
and {\fmvrc the} event-by-event \dt\ uncertainty for \Bzb(\Bz) tags. 
The shaded area shows the background contribution to the distributions. 
\label{fig:dtbcp}}
\end{figure}

Figs.~\ref{fig:dtbflav} and \ref{fig:dtbcp} show the \dt\
distributions of {\fmvrc events confined to the signal region, defined as}
$\mes>5.27$ \gevcc\ for the
\bflav\ and \bcpks\ samples, and $| \Delta E | < 10$ \mev\ for the \bcpkl\ sample. The
points correspond to data. The curves 
correspond to the projections of the likelihood fit allowing for \CPT\ violation,
weighted by the appropriate relative amounts of signal and
background. The background contribution is indicated {\fmvcw by} the shaded area.

%% file: mistag-table.tex
\setlength{\extrarowheight}{3pt}
\begin{tabular}{l|c|c|c|c}
\hline
Tagging            & $\tau^\alpha(\%)$ & $\wasig_0(\%)$ & $\wasig_{\rm slope}$ & $\dwasig(\%)$ \\
category           &                &          &                  & \\
\hline \hline
{\tt Lepton}    & $\phan 9.4  \pm  0.2$ & $\phan 2.6\pm0.7$ & $0$ (fixed) & $-1.2 \pm 1.2$  \\
{\tt Kaon I}    & $17.2 \pm  0.3$ & $\phan 2.0\pm2.0$ & $0.13\pm0.04$ & $-2.7 \pm 1.3$  \\
{\tt Kaon II}   & $19.9 \pm  0.3$ & $15.9\pm2.4$ & $0.07\pm0.04$ & $-4.2 \pm 1.3$  \\
{\tt Inclusive} & $19.9 \pm  0.3$ & $26.5\pm2.5$ & $0.07\pm0.04$ & $-2.9 \pm 1.3$  \\
{\tt UnTagged}  & $33.6 \pm  0.6$ & $50$ (fixed) & $0$ (fixed) & $0$ (fixed)  \\
\hline
\end{tabular} 

%% file: resolution-table.tex
\setlength{\extrarowheight}{3pt}
\begin{tabular}{l|c||c|c}
\hline
Parameter & Fitted value & Parameter & Fitted value\\ 
\hline \hline
$S_{\rm core}$                          & $\pham 1.25 \pm 0.04$ & 
$S_{\rm tail}$                          & $\pham 5.7\pm0.8$  \\
$\delta_{\rm core}^{\tt Lepton}$        & $\pham 0.02 \pm 0.07$ &
$\delta_{\rm tail}$                     & $-1.5 \pm 0.5$ \\
$\delta_{\rm core}^{\tt Kaon I}$        & $-0.27 \pm 0.05$ &
$f_{\rm tail}$                          & $0.034 \pm 0.010$ \\
$\delta_{\rm core}^{\tt Kaon II}$       & $-0.32 \pm 0.04$ &
$\sigma_{\rm out}$                      & 8~ps (fixed)  \\
$\delta_{\rm core}^{\tt Inclusive}$     & $-0.30 \pm 0.04$ &
$\delta_{\rm out}$                      & 0~ps (fixed) \\
$\delta_{\rm core}^{\tt UnTagged}$      & $-0.28 \pm 0.03$  &  
$f_{\rm out}$                           & $0.0003 \pm 0.0012$ \\
\hline
\end{tabular} 

%% file: munus-table.tex
\setlength{\extrarowheight}{3pt}
\begin{tabular}{l|c|c}
\hline
Parameter       & Nominal fit & Counting-based method \\
\hline\hline
$\nu^{\rm sig}$               & $\pham 0.011 \pm 0.008$ & $\pham 0.007\pm0.008$ \\
$\mu^{\tt Lepton,\rm sig}$    & $\pham 0.024 \pm 0.022$ & $\pham 0.029\pm0.042$ \\
$\mu^{\tt Kaon I,\rm sig}$    & $-0.022 \pm 0.017$      & $-0.022\pm0.029$ \\
$\mu^{\tt Kaon II,\rm sig}$   & $\pham 0.014 \pm 0.016$ & $\pham 0.004\pm0.027$ \\
$\mu^{\tt Inclusive,\rm sig}$ & $\pham 0.014 \pm 0.016$ & $\pham 0.025\pm0.027$ \\
\hline
\end{tabular} 

%% file: physpar-table.tex
\setlength{\extrarowheight}{3pt}
\begin{tabular}{l|c|c}
\hline
Parameter & Fit with $\z$ free & Fit with $\z=0$ \\ 
\hline \hline 
$\sgndGoverG$       &  $ -0.008 \pm 0.037 $ & $ -0.009 \pm 0.037 $     \\
$\absqop$         &  $\pham 1.029 \pm 0.013 $ & $\pham 1.029 \pm 0.013 $      \\
$\reZparflat$  &  $\pham 0.014 \pm 0.035 $ & $-$ \\
$\imZ$            &  $\pham 0.038 \pm 0.029 $ & $-$ \\
\hline
$\dM\ (\ps^{-1})$              & $0.521\pm0.008$   & $0.521\pm0.008$ \\
$\imlambcpflat$       &  $0.752\pm0.067$  & $0.741\pm0.067$ \\
\hline 
$\imlambtagflat$      &  $\pham 1.5 \pm 1.2 $ & $\pham 0.5 \pm 1.0 $ \\
$\imlambbartagflat$   &  $ -0.1 \pm 1.2 $ & $\pham 0.8 \pm 1.0 $ \\
$\imlambflavflat$     &  $\pham 2.3 \pm 1.1 $ & $\pham 1.4 \pm 0.9 $ \\
$\imlambbarflavflat $ &  $-0.6 \pm 1.1 $ & $\pham 0.1 \pm 0.9 $ \\
\hline
\end{tabular}

%% file: correlation-table.tex
\setlength{\extrarowheight}{3pt}
\begin{tabular}{l|l|r|r} 
\hline
Parameter & Parameter  & \multicolumn{2}{c}{Correlation (\%)} \\ \cline{3-4}
          &            & \phantom{d} $\z$ free      &  $\z=0$ \\
\hline \hline
\dM  & \sgndGoverG & $-1.3$ & $-0.9$ \\
     & \absqop & $-2.8$ & $-2.8$ \\
     & \imlambcpflat & $-5.6$ & $-5.3$ \\
     & \reZparflat & $7.0$ & $-$ \\
     & \imZ       & $-0.2$ & $-$ \\
\hline
\sgndGoverG & \absqop & $11.0$ & $10.8$  \\
            & \imlambcpflat & $0.4$ & $0.2$ \\
            & \reZparflat & $-7.9$ & $-$ \\
            & \imZ       & $-1.8$ & $-$ \\
\hline
\absqop & \imlambcpflat & $-1.0$ & $-1.5$ \\
     & \reZparflat & $-2.4$ & $-$ \\
     & \imZ       & $-1.1$ & $-$ \\
\hline
\imlambcpflat & \reZparflat & $-10.9$ & $-$ \\
     & \imZ       & $17.4$ & $-$ \\
\hline
\reZparflat & \imZ & $-3.4$ & $-$ \\
\hline
\end{tabular}

%% file: correlation-table-top.tex
\setlength{\extrarowheight}{3pt}
\begin{tabular}{l|l|r} 
\hline
Physics parameter & Parameter  & Correlation (\%) \\
\hline \hline
\dM  & $w^{\tt Lepton,\rm sig}_0$ & $-20.1$ \\
     & $f_{\tt tail}$     & $18.7$ \\
     & $S_{\tt tail}$     & $-15.4$ \\
\hline
\sgndGoverG & \absqop & $11.0$ \\
\hline
\absqop & $\nu^{\tt sig}$       & $65.1$ \\
        & $\dw^{\tt KaonII,\rm sig}$    & $-22.5$ \\
        & $\mu^{{\tt Lepton,sig}}$  & $22.4$ \\
        & $\dw^{\tt KaonI,\rm sig}$     & $-22.4$ \\
        & $\dw^{\tt Inclusive,\rm sig}$ & $-15.5$  \\
        & $\mu^{{\tt KaonI,sig}}$   & $13.9$ \\
        & $\dw^{\tt Lepton,\rm sig}$    & $-13.5$ \\
        & \sgndGoverG           & $11.0$ \\
\hline
\imlambcpflat & \imZ            & $17.4$\\
          & \imlambtagflat      & $14.4$ \\
          & \imlambflavflat     & $13.6$ \\
          & \reZ                & $-10.9$ \\
\hline
\reZparflat    & \imlambcpflat  & $-10.9$ \\
\hline
\imZ & \imlambtagflat      & $61.6$ \\
     & \imlambflavflat     & $57.7$  \\
     & \imlambbartagflat   & $-56.6$  \\
     & \imlambbarflavflat  & $-54.0$  \\
     & \imlambcpflat       & $17.4$ \\
     & $\nu^{\tt sig}$     & $11.0$ \\
\hline
\end{tabular}

%% file: checks.tex
We use data and Monte Carlo samples to perform validation studies of
the analysis technique.  {\rncnov The} Monte Carlo tests include
studies with parameterized fast Monte Carlo as well as full GEANT4-simulated samples. 
Checks with data are performed
with control {\fmvcw samples, where} no \dG\ and \CP-, \T-, and \CPT-violating effects are
expected.  Other checks are made by analyzing the actual data sample,
but using alternative tagging, vertexing, and fitting configurations.

\subsection{Monte Carlo Simulation Studies}
\label{sec:checks-mc}

{\fmvrc A test} of the fitting procedure is performed {\fmvrc with}
parameterized Monte Carlo simulations {\fmvrc consisting of} 300 experiments
{\fmvrc generated} with a sample size and composition corresponding to that
of the data. The mistag
probabilities and \dt\ distributions are generated according to the model used
in the likelihood function. 
{\fmvcw The physics parameters are generated according to the values found
in the data \footnote{Studies with alternative values in a wide range of variation 
have also been performed.}.}
The nominal fit is then performed on each
of these experiments.  Each experiment uses the set 
of \mes\ ($\Delta E$) and $\sdt$ values observed in the non-\KL(\KL) sample.  The
\rms\ spread of the residual distributions for all physics parameters
(where the residual is defined as the difference between the fitted
and generated values) is found to be consistent, within 10\%, with
the mean (Gaussian) statistical errors reported by the fits. Moreover,
{\fmvrc it has been verified using these experiments}
that the asymmetric 68\% and
90\% confidence-level intervals {\fmvrc obtained from the fits} provide the correct {\fmvrc statistical} coverage.  

In all cases, the mean
values of the residual distributions are consistent with
no measurement bias. A systematic error due to the limited precision of this
study is assigned to each physics parameter.  The statistical errors
on all the physics parameters (Table \ref{tab:phys-param}) and the
calculated correlation coefficients among them (Tables
\ref{tab:correlation} and \ref{tab:correlation-top}), extracted from 
the fit are consistent with the range of values obtained from these experiments. We
find that 24\% of the fits result in a value of the
log-likelihood that is greater (better) than that found in data.

In addition, {\fmvrc samples} of signal and background Monte Carlo events
generated with a 
full
detector simulation are used
to validate the measurement. The largest samples are generated with
{\fmvrc \dGoverG, \mbox{$\absqop-1$}, and \z\ all equal to zero}, but additional samples
are also produced with relatively large values of these parameters. Other values
(including those measured in the data) are generated with reweighting
techniques. The signal Monte Carlo {\rncnov events} are split into
samples whose size and proportions of \bflav,
\bcpks, and \bcpkl\  are similar to those of the actual data set. 
To check whether
the selection criteria or the analysis and fitting procedures
introduce any bias in the measurements, the fit (to signal alone)
is then carried out on these experiments, allowing for \CPT\ violation. 
The small combinatorial
background in these signal samples is suppressed  
by restricting the fit to {\fmvcw the} events in
the signal region.  
Fits to a sample without background, using the
true \dt\ distribution and true tagging information, are also
performed.  The means of the residual distributions from all these
experiments for all the physics parameters are consistent with zero,
confirming that there is no measurement bias. The \rms\ spreads are
consistent with the average reported errors.
A systematic error is assigned to each
physics parameter corresponding to the limited Monte Carlo statistics 
for this test. 

The effect of backgrounds is evaluated by adding an
appropriate fraction of background events to the signal Monte Carlo
sample and performing the fit. The \bcp\ background samples are
obtained either from simulated $\B \to \jpsi X$ events or $\Delta E$
sidebands in data, while the \bflav\ backgrounds are obtained from
generic $\B\Bb$ Monte Carlo. We find no evidence for bias in any of the
physics parameters.

\subsection{Cross-checks with Data}
\label{sec:checks-data}

We fit subsamples defined by
tagging category or %running
{\fmvcw data taking} 
period. Fits using only the 
$\Bz \to D^{(*)-} X^+$ or $\Bz \to \jpsi \Kstarz(\Kp\pim)$ 
channels for \bflav, and only \bcpks\ or only \bcpkl\ for \bcp\ are also
performed. We find no statistically significant differences in the results
for the different subsets.  We also vary the maximum allowed values of $| \dt |$  between 5 and 30\ps,
and of $\sdt$ between 0.6 and 2.2\ps.  Again, we do not find statistically significant changes
{in the physics parameters.}

In order to verify that the results are stable under variation of the
vertex algorithm used in the measurement of \dt, we use
alternative (less powerful) methods, {\fmvrc described in Sec.~VIII.C.5 in Ref.}~\cite{ref:babar-stwob-prd}. 
To reduce statistical fluctuations due to different events being
selected, the comparison between the alternative and nominal methods
is performed using only the {events accepted by both methods.}
Observed variations are {\ffsix small compared} with
the systematic error assigned to the resolution
function (see Sec.~\ref{sec:systematics}).

The stability of the results under variation of the tagging algorithm
is studied by repeating the fit using the tagging 
algorithm {\fmvrc described in Sec.~IV in Ref.}~\cite{ref:babar-stwob-prd}. 
The algorithm used in that analysis
has an effective tagging efficiency $Q=\sum_{\alpha}
\tau^{\alpha}\left(1-2\wa \right)^2$ about 7\% lower than the one used here. 
The variations observed in the physics parameters {\fmvcw are} consistent with
the statistical differences.

The average \Bz\ lifetime is fixed in the nominal fit to the PDG value~\cite{ref:pdg2002}. 
This value is obtained by averaging
measurements based on flavor-eigenstate samples and by assuming
negligible effects from \dGoverG, \absqop, and \z. 
Measurements that do not use tagged events are 
{\fmvsix largely insensitive to}
\absqop\ and \z, but would be {\fmvsix affected}, at second order,  by a non-zero value
of \dGoverG, as discussed 
in Sec.~\ref{sec:decayrates}. Therefore we do not expect {\fmvsix sizeable}
effects from the fixed average \Bz\ lifetime.  However, to check the
consistency of the result, the fit is repeated with the average lifetime 
left free. The resulting \taub\ is about {\ffsix one standard deviation}
below the nominal value assumed in our analysis, 
taking into account the statistical
error from the fit and the present \taub\ uncertainty. 
\remark{deleted sentence}
As described in Sec.~\ref{sec:sys.C}, a systematic
error is assigned using the variation of each physics parameter when
the fit is repeated 
{\fmvrc with \taub\ fixed to the value obtained when it is floated, which corresponds to} 
a change of about twice the present PDG error ($\pm 0.032$\ps).

Similarly, fits with \Rcp\ free have been performed. The resulting \Rcp\
value is consistent with unity (the fixed nominal value) within one
standard deviation (statistical only).  
\remark{deleted sentence}
As described in Sec.~\ref{sec:sys.C}, 
systematic errors due to fixing \Rcp\ at unity are set by changing
\Rcp\ by twice the statistical uncertainty determined by leaving it free in the fit
($\pm 10\%$).  The resulting variation in each parameter is taken as
the systematic error.

The robustness of the fit is also tested by modifying the nominal PDF
normalization, as described by Eq. (\ref{eq:norm2}), so that the
analysis is insensitive to the relative {\fmvsix number} of \Bz\ and \Bzb\ tagged events. 
As a consequence, the statistical {\ffsix error on} \absqop\ is
dramatically {\ffsix increased}, since the sensitivity to this parameter comes
largely from the differences in time-integrated \Bz\ and \Bzb\ rates.
In addition, the fit is also performed assuming an independent set of
resolution function parameters for each tagging category. In all cases
the results are {\ffsix consistent} 
with the nominal fit results. 
Finally, the tagging efficiencies $\tau^{\alpha}$
are alternatively determined 
{\fmvsix for each sample (\bflav, \bcpks\ and \bcpkl) separately, rather than
using a common estimate from the \bflav\ sample, as in the nominal fit.}
The changes in the values
of the physics parameters are negligible.

Control samples in data from \Bu\ decays (treated in a way analogous 
to that described in 
Sec.~\ref{sec:sample}) are also used to validate the analysis
technique, since in these samples we expect zero values for \dGoverG,
\mbox{$\absqop-1$} and $\z$. For the \bflav\ sample we use the 
$\Bu \to \Dzb\pip, \Dstarzb\pip$ decay channels, 
and for the \bcp\ sample the
{\fmvsix decays of charged-\B\ mesons to charmonium plus a charged $K$ or $K^*$ (see Table~\ref{table:modes}).}
The check is performed {\fmvcw by} fixing $\dM=0$ and
\mbox{$\absqop=1$} in the \bflav\ sample, and 
{\fmvcw assuming maximal mixing} (\mbox{$\dM=0.489${~\rm ps}$^{-1}$} \cite{ref:pdg2002}) 
in the \bcp\ sample, {\fmvcw and} fitting for \imlambcpflat, \sgndGoverG,
\reZparflat\ and \imZ.  No statistically significant deviations from
zero are observed.

%% file: systematics.tex
We estimate systematic uncertainties with
studies performed on both data and Monte Carlo simulation samples. A
summary of the  sources of non-negligible uncertainties is shown in Table
\ref{tab:systglob}.  In the following, the individual 
contributions are referenced by the lettered lines in the Table.

\begin{table*}[!ht]
\caption{Summary of systematic uncertainties on the measurements of \sgndGoverG, 
\absqop, \reZparflat, and \imZ.}
\centering
\input{systematics-cpt-table}

\label{tab:systglob}
\end{table*}

\subsection{Likelihood Fit Procedure}

Several sources of systematic uncertainties due to the likelihood fit procedure
are considered. We  include the results from the tests
performed using the {\fmvrc parameterized} Monte Carlo sample (a) and the full
GEANT4 signal Monte Carlo sample (b), as described in
Sec.~\ref{sec:checks-mc}.  
No statistically significant bias (mean of the residual distributions) is
observed. Thus, we assign a systematic error equal to the
statistical uncertainty on the bias.
No corrections are applied to the central
values extracted from the fit to the data. 
Note that the GEANT4 contribution
accounts for residual differences {\fmvsix between the} \bflav, \bcpks, and \bcpkl\ samples
in the mistag {\ffsix probability},
resolution function, and $\nu$ and $\mu^\alpha$ parameters. 
{\fmvsix It also includes residual} differences in \dt\ resolution for
correct and wrong tags. 

We also consider the impact on the measured physics parameters of
{\rncnov normalizing the time-dependent PDFs to the full interval
$-\infty <\dt<\infty$.} The effect is evaluated 
{\fmvsix by repeating the fit}
using a normalization in the range defined by the \dt\
cut. Finally, the fixed tagging efficiencies are varied within their
statistical uncertainties.  The two contributions are negligible.

\subsection{\dtResolutionFunction}

The \dt\ resolution model used in the analysis, consisting of the sum of
three Gaussian distributions, is expected to be flexible enough to
represent the experimental resolutions. To assign a systematic
error for this assumption we use the alternative model
described in Sec.~\ref{sec:decaytime}, with a Gaussian distribution
plus the same Gaussian convolved with one exponential function, for
both signal and background.  The results for all physics parameters
obtained from the two resolution models are consistent and we assign 
the difference of central values as a systematic uncertainty (c).

In addition, a number of parameters that are {\fmvsix inherent} to the
determination of \dt\ are varied according to known uncertainties. The
PEP-II boost, estimated from the beam energies, has an uncertainty of
0.1\%~\cite{ref:babar-nim}. The absolute $z$-scale uncertainty is
evaluated to be less than 0.4\%. This estimate is obtained by
measuring the beam pipe dimensions with scattered protons and
comparing to optical survey data. Therefore, the boost and $z$-scale
{\fmvsix systematic uncertainties} are evaluated {\fmvcw conservatively} by 
varying by $\pm0.6\%$ the reconstructed
\dt\ and $\sdt$ (d).  
{\fmvrc As the beam spot is much smaller in the vertical than
in the horizontal dimension, its vertical position and size 
is more relevant in the vertex fits. Hence}
the uncertainty on the position and size of the beam spot used in the vertex fits is
{\fmvrc taken} into account by changing the vertical position by up to 40\mum
and increasing the vertical size from 10\mum to 60\mum (e).
Finally, the systematic uncertainty due to possible SVT internal
misalignment is evaluated by applying a number of possible misalignment
scenarios to a sample of simulated events and comparing the values of
the fitted physics parameters from these samples to the case of
perfect alignment (f).

Fixing the width and bias of the outlier Gaussian {\fmvrc distribution} 
in the resolution function to 8\ps and 0\ps, respectively,
{\fmvsix is a potential source of bias}.
To estimate the {\fmvrc corresponding systematic}
uncertainty we add in quadrature the
variation observed in the physics parameters when the bias changes
by $\pm$5\ps, the width varies between 6\ps and 12\ps, and the outlier
distribution is assumed to be flat (g).

\subsection{Signal Properties}
\label{sec:sys.C}

As described in Sec.~\ref{sec:checks-data}, the uncertainty from fixing the 
average \Bz\ lifetime is evaluated by changing its central value by  $(\pm 0.032\ps)^{-1}$ (h), 
twice the PDG error~\cite{ref:pdg2002}. Possible direct \CP\ violation in the \bcp\ sample
is taken into account by varying \Rcp\ by $\pm 10\%$ (i).

{\fmvsix Systematic uncertainties related to \dckm\ decays arise because we fix the
real parts of \lambtagbare, \lambbartagbarbare, \lambflavbare, and \lambbarflavbarbare\ to zero.}
In order to evaluate this contribution, {\fmvsix we generate} samples of parameterized Monte
Carlo samples tuned to the data sample, {\fmvrc scanning the \dckm-decay phases over their full allowed
range (0-2$\pi$) and}
assuming a single
hadronic decay channel contributing to the \btag\ and to the \bflav.
{\ffsix Samples are generated with values of 
$|\overline{A}_{\tagv}/ A_{\tagv} |$
and 
$| \overline{A}_{\flavv}/ A_{\flavv}  |$
equal to 0 and 0.04, corresponding to 100\% variation of the value 0.02 used in the nominal fit.}
For the {\tt Lepton} tagging category, 
dominated by semileptonic \B\ decays, we assume $\lambtagbare$ to be zero. 
{While the ratio of CKM matrix elements leads to the nominal value  
$|\overline{A}_{\tagv}/ A_{\tagv}|=|\overline{A}_{\flavv}/ A_{\flavv}|=0.02$, this is not a reliable estimate for any
single decay mode.  Examination of the {\fmvrc \dckm\ charmed-meson decay}
$D^0\to K^+\pi^-$ shows good agreement with expectations from CKM matrix elements, albeit with large uncertainties, 
but the singly-\ckm-suppressed decays $D^0\to \pi^+\pi^-$ and $D^0\to K^+K^-$
show deviations as large as a factor of two.  However, %traditional arguments suggest that 
{\fmvrc when we} 
sum over many channels, 
as we do here both for tagging states and for flavor eigenstates, quark-level predictions are much more
reliable than they are for a single channel.  Allowing for 100\% variation from
the nominal value of 0.02 is thus conservative.}
}

Using the fit results from all these samples, we determine 
{\fmvcw the offsets} with respect to the generated
value and its statistical uncertainty, for {\fmvrc a complete 
sampling} of {\fmvrc \dckm-decay} phases. 
The systematic error assigned is the largest value
among all configurations (j).  This is the dominant source of
systematic uncertainty for the measurement of \reZparflat\ {\fmvrc and is due primarily
to the influence of \dckm\ decays} in the tagging-\B\ meson. 
{\fmvsix The effect of using a single effective channel for the flavor and all tagging category states
has been estimated by splitting the \bflav\ and \btag\ samples generated with the parameterized Monte Carlo
into equally sized subsamples. For the different combinations of {\fmvrc \dckm-decay} phases, the observed offset is
about the average of the biases obtained using the single effective channel.
{\fmvrc Therefore,} the largest offset among all
configurations is smaller than that observed for a single channel.
This shows that our prescription to describe the {\fmvrc effects from \dckm\ decays} and to
assign the systematic uncertainties assuming a single effective
channel is conservative.}

Charge asymmetries induced by a difference in the detector response
for positive and negative tracks are included in the PDF and extracted
together with the other parameters from the time-dependent
analysis. Thus, they do not contribute to the systematic error, but
rather are incorporated into the statistical error at a level
determined by the size of the \bflav\ data sample. Nevertheless, in
order to account for any possible residual effect, we assign a
systematic uncertainty as follows.  We rerun the \B\ reconstruction,
{\fmvsix vertex-finding, and tagging algorithms} after removing randomly and uniformly (no
momentum or angular dependence) 5\% of positive and, separately, negative
tracks in the %large statistics dedicated
full Monte Carlo sample.
This value of 5\% is on average more than a factor of three larger than the
precision with which the parameters $\nu^{\rm sig}$ and $\mu^{\alpha,\rm sig}$ have been
measured in the data.  Half the difference between the results
obtained for positive and negative tracks is assigned as a systematic error
(k).

\subsection{Background Properties}

The event-by-event signal probability 
$p^{\alpha}_{\reci}(\mes)$ for \bflav\ and \bcpks\ samples is fixed to the
values obtained from the \mes\ fits. We compare the results from
the nominal fit to the values obtained by varying  all the \mes\ distribution 
parameters by $\pm 1\sigma$, taking into account their
correlations. This is performed simultaneously for all tagging
categories, and independently for the \bflav\ and \bcpks\
samples. {As an alternative}, we also use a flat signal probability
distribution: events belonging to the sideband region (\mes$<$5.27
\gevcc) are assigned a signal probability of zero, while we give a
signal probability equal to the purity of the corresponding sample to
signal region events (\mes$>$5.27 \gevcc). The differences among
fitted physical parameters with respect to the default method are
found to be consistent. 
We determine the systematic error due to this parameterization by varying the 
signal probability by 
its statistical error.
The final systematic error is taken to be the larger of the one-sigma variations
found for the two methods (l). The uncertainty on the fraction of
peaking background is estimated by varying the fractions according to
their uncertainties separately for the \bflav\ sample and each \bcpks\
decay mode (m). The effective $\eta_{\CPscript}$ of the
\bcpks\ peaking background, assumed to be zero 
in the nominal fit, is also varied between $+1$ and $-1$
and the variations induced are negligible.

Another source of systematic uncertainty 
is the assumption that the \dt\ behavior of the combinatorial
background in the \mes\ sideband region is {the same as it is} in the signal region. 
However, the background composition
changes {gradually as a function {\fmvcw of \mes, since} the fraction due to
continuum production slowly decreases as \mes\ increases}. 
{\fmvrc To study the impact of variable \dt\ behavior over the \mes\ range,}
we vary the lower edge of the \mes\ distributions
from 5.20\gevcc\ to 5.27\gevcc, simultaneously for the
\bflav\ and \bcpks\ samples, observing  good stability in the results. 
We also split the sideband region in seven equal slices each 10
\mevcc\ wide and 
{\fmvcw repeat the fit in each of these slices.}
The results obtained for all physics parameters
and \mes\ slices are then linearly extrapolated to the \B-mass signal region.
The quadratic sum of the extrapolation
and the error on it is assigned as a systematic uncertainty (n).

As described in Sec.~\ref{sec:method}, the likelihood fit
assumes that there are no effects of \dG, \CP, \T\ or \CPT\ violation, mixing, and
\dckm\ decays in the combinatorial background
components (\bflav\ and \bcpks\ samples) and in the non-\jpsi\
background (\bcpkl\ sample). To evaluate the effect of this {\fmvcw assumption, we} repeat 
the fit assuming for the background non-zero values of \dG, \mbox{$\absqop-1$},
\z, \imlambcpflat, and \dM, and varying $\eta_{\CPscript}$ of the background by
$\pm 1$.  The check is performed by introducing an
independent set of physics parameters {\fmvcw in the PDF} and assuming maximal mixing and
\CP\ violation (\dM\ and \imlambcpflat\ fixed to $0.489\ps^{-1}$~\cite{ref:pdg2002} and 0.75~\cite{ref:sin2b-babar},
respectively). {\fmvrc \dckm-decay} effects are included by
assuming the maximal values (0.04) of $| \overline{A}_{\tagv}/A_{\tagv} |$ 
and $| \overline{A}_{\flavv} /A_{\flavv} |$, and scanning all the
possible values of the \Bz\ and \Bzb\ phases for \bflav\ and
\btag. The systematic uncertainty is evaluated simultaneously for all
these sources (o).

The systematic errors due to the \Bu\ decay rate 
are evaluated by {varying its value by the PDG uncertainty~\cite{ref:pdg2002}}. 
{The effect is negligible.} The \Bu\ mistags
and the differences in the fraction of \Bu\ and \Bub\ mesons that are
tagged and reconstructed are varied according to their statistical
errors as obtained from the fit to the \Bu\ data. These errors
are found to be negligible.

Uncertainties from charge asymmetries in
combinatorial background components (neglected in the nominal fit)
are evaluated by repeating the fit with a new set of $\nu$ and
$\mu^\alpha$ parameters. The measured values of $\nu$ and $\mu^\alpha$
are found to be compatible with zero and the variation of the physical
parameters with respect to the nominal fit is assigned as a systematic
error (p).

For the \bcpkl\ channel, %~\cite{ref:babar-stwob-prd}, 
the signal and
non-\jpsi\ background fractions are varied according to their
statistical uncertainties, obtained from the fit to the \de\
distribution.  We also vary background parameters, including the
$\jpsi X$ branching fractions, the assumed $\eta_{\CPscript}$, the \de\ shape,
and the fraction and effective lifetime of the prompt and non-prompt
non-\jpsi\ components. The differences observed between data and Monte
Carlo simulation for the \KL\ angular resolution and for the fractions
 of $\Bz \to \jpsi\KL$ events reconstructed in the EMC and IFR
are used to evaluate a systematic uncertainty due to the simulation of the \KL\
reconstruction.
Finally, an additional contribution is assigned to the correction
applied to {\tt Lepton} events due to the observed differences in
flavor tagging efficiencies in the \jpsi\ sideband relative to \bflav\
and inclusive \jpsi\ Monte Carlo samples.  Conservatively, this error is
evaluated {\fmvsix by} comparing the fit results with and without the correction.
The total \bcpkl-specific systematic error is evaluated by taking the
quadratic sum of the individual contributions (q).

\subsection{Summary of Systematic Uncertainties}

All individual systematic contributions described above and summarized
in Table \ref{tab:systglob} are added in quadrature. The dominant
source of systematic error in the measurement of \reZparflat\ is due
to our limited knowledge of the \dckm\ decays, which
also contributes significantly to the {\ffsix uncertainties} on the  other measurements. The limited
Monte Carlo {\ffsix sample size is} a dominant source of systematic error for
\absqop, \imZ,  and to a lesser extent for
\sgndGoverG. Residual charge asymmetries, {\fmvsix mainly due to limited simulation statistics}, 
dominate the systematic error on \absqop.
Our limited knowledge of the beam spot and SVT alignment reflects
significantly on \imZ\ and \sgndGoverG.  
The systematic error on
\sgndGoverG\  receives a non-negligible contribution from our
{\rncnov incomplete} understanding of the resolution function. 

The systematic uncertainties 
on \sgndGoverG\ and \absqop\ when \CPT\ {\fmvcw invariance} is assumed %!!!!!!!!! 
are evaluated similarly, and found to be
consistent, within the statistical fluctuations of the Monte Carlo
simulation, with those found for the analysis when \CPT\ violation is allowed.

%% file: systematics-cpt-table.tex
\renewcommand{\arraystretch}{1.3}
\begin{tabular}{l|c|c|c|c}
\hline
Systematics source & \sgndGoverG & \absqop & \reZparflat & \imZ \\ 
\hline\hline
\multicolumn{5}{c}{Likelihood fit procedure} \\
\hline
(a) Parameterized MC test & 
$\phal 0.003$ & $\phal 0.001$ & $\phal 0.003$ & $\phal 0.003$ \\
(b) GEANT4-simulation test &
$\phal 0.005$ & $\phal 0.007$ & $\phal 0.004$ & $\phal 0.016$ \\
\hline \hline
\multicolumn{5}{c}{\dtresolutionfunction} \\
\hline
(c) Resolution function parameterization  &
$\phal 0.007$ & $\phal 0.001$ & $\phal 0.008$ & $\phal 0.003$ \\
(d) $z$ scale and boost  &
$\phal 0.003$ & $\phal 0.001$ & $\phal 0.002$ & $< 0.001$ \\
(e) Beam spot  & 
$\phal 0.008$ & $\phal 0.002$ & $\phal 0.001$ & $\phal 0.011$ \\
(f) SVT alignment  &
$\phal 0.006$ & $\phal 0.001$ & $\phal 0.001$ & $\phal 0.011$ \\
(g) Outliers   &
$\phal 0.002$ & $< 0.001$ & $< 0.001$ & $< 0.001$ \\
\hline \hline
\multicolumn{5}{c}{Signal properties} \\
\hline
(h) Average \Bz\ lifetime  &
$\phal 0.004$ & $\phal 0.001$ & $\phal 0.004$ & $< 0.001$ \\
(i) Direct \CP\ violation  &
$\phal 0.002$ & $\phal 0.004$ & $\phal 0.001$ & $\phal 0.003 $ \\
(j) \dckm\ decays  &
$\phal 0.008$ & $\phal 0.004$ & $\phal 0.032$ & $\phal 0.006 $ \\
(k) Residual charge asymmetries  &
$\phal 0.005$ & $\phal 0.006$ & $\phal 0.004$ & $\phal 0.006$ \\
\hline \hline
\multicolumn{5}{c}{Background properties} \\
\hline
(l) Signal probability  & 
$\phal 0.002$ & $\phal 0.001$ & $\phal 0.002$ & $\phal 0.001$ \\
(m) Fraction of peaking background  &
$< 0.001$ & $< 0.001$ & $\phal 0.004$ & $< 0.001$ \\
(n) $\dt$ structure  &
$\phal 0.002$ & $\phal 0.001$ & $\phal 0.001$ & $\phal 0.001$ \\
(o) \dG/\CP/\T/\CPT/Mixing/\dckm\ content &
$\phal 0.001$ & $\phal 0.002$ & $\phal 0.002$ & $< 0.001$ \\ 
(p) Residual charge asymmetry  &
$< 0.001$ & $\phal 0.001$ & $< 0.001$ & $< 0.001$ \\ 
(q) \KL-specific systematic errors &
$\phal 0.004$ & $< 0.001$ & $\phal 0.004$ & $\phal 0.003$ \\
\hline\hline
Total systematic uncertainties & $\phal 0.018$    & $\phal 0.011$ & $\phal 0.034$ & $\phal 0.025$ \\ 
\hline
\end{tabular}
\renewcommand{\arraystretch}{1.0}

%% file: summary.tex
The conventional analyses of mixing and \CP\ violation in the 
neutral-\B-meson system neglect possible contributions from several 
sources that are expected to be small.  These include the difference of
the decay rates of the two neutral-\B-meson mass eigenstates, 
the \CP- and \T-violating quantity $\absqop-1$, 
and potential \CPT\ violation.  To measure or extract limits on these quantities
requires the full expressions for time dependence in mixing and 
\CP\ violation and consideration of systematic {\fmvsix effects} that might mimic
{\fmvrc the} fundamental asymmetries we seek to measure.
{\fmvrc Such systematic effects could be induced by} 
detector charge asymmetries, different resolution functions for positive and negative
\dt, and \dckm\ decays for both fully reconstructed
final flavor states and \nonleptonic\ tagging states.

{\fmvsix A limit on the decay-rate difference of $|\dG/\G| <80\%$ at 95\% confidence-level
was obtained by CLEO~\cite{ref:cleochid} using the 
time-integrated mixing parameter $\chi_d$ and the mass difference \dM\ extracted
under the assumption $\dG=0$. Using \Z\ decays, DELPHI~\cite{ref:delphi} has recently performed
a time-dependent study of semileptonic \B\ decays inclusively reconstructed. Assuming
no \CP, \T, or \CPT\ violation in mixing, they quote the limit $| \dG/\G | <18\%$ at 95\% confidence-level.
}

Both \absqop\ and \imZ\ were measured by OPAL~\cite{ref:otherCPTBtests}, using \Z\ decays to $b{\overline
b}$ pairs and assuming $\dG=0$. Neutral-\B-meson oscillations were studied by observing a single
lepton indicative of a $B$ decay and the jet-charge associated with
both the jet containing the lepton and the other jet. Because
the multiparticle final states provide essentially uncorrelated \B\ mesons, 
the issue of \dckm\ decays is obviated.  The results were $\re\,
\epsilon_B=0.006\pm 0.010 \pm 0.006$, equivalent to
$|q/p|=0.988\pm0.020\pm0.012$, and 
$\im \epsilon_B=-0.020\pm 0.016\pm 0.006$, equivalent to 
$\imZ=0.040\pm0.032\pm0.012$.  
\fmvrf{Combining the earlier \absqop\ measurements, all obtained 
assuming $\dG=0$, gives $\absqop=0.9993\pm0.0064$~\cite{ref:pdg2003}.}
Belle has used
dilepton events to obtain limits on \CPT\ violation~\cite{ref:belle_dilepton}.  
Assuming {\fmvcw that} $\Delta\Gamma=0$ and
that \CP\ violation in mixing can be ignored, they find
$\re\cos\theta=-\reZ=0.00\pm0.12\pm0.02$ and 
$\im\cos\theta=-\imZ=0.03\pm 0.01\pm 0.03$.

Our analysis of approximately 31,000 fully reconstructed flavor
eigenstates and 2,600 \CP\ eigenstates sets new limits on the difference
of decay rates of \Bz\ mesons, and on the \CP, \T, and \CPT\ violation
intrinsic to $\BzBzb$ mixing. The six independent parameters
governing oscillations \mbox{(\dM, \dGoverG)}, \CPT\ and \CP\ violation \mbox{(\reZ, \imZ)}, and
\CP\ and \T\ violation (\imlambcpbare, \absqop) are extracted from a single fit
of both fully reconstructed \CP\ and {\fmvsix flavor} events, tagged and
untagged.  This provides the sensitivity required to separate 
{\fmvsix all}
effects we seek from asymmetries in detector response and from
potentially obscuring correlations in the decays of the two \B\ mesons. 
The results are
\newcommand{\resline}[5]{\quad #1 \pm #2\mathrm{(stat.)}  \pm #3\mathrm{(syst.)} [#4, #5]}
$$\begin{array}{l}
\sgndGoverG = \\
\resline{-0.008}{0.037}{0.018}{-0.084}{0.068}~,\\
\absqop = \\
\resline{\phantom{+}1.029}{0.013}{0.011}{\phantom{+}1.001}{1.057}~,\\
\reZparflat = \\
\resline{\phantom{+}0.014}{0.035}{0.034}{-0.072}{0.101}~,\\
\imZ = \\
\resline{\phantom{+}0.038}{0.029}{0.025}{-0.028}{0.104}~.\\
\end{array}$$

\begin{figure}[!h]
\centering
\epsfig{file=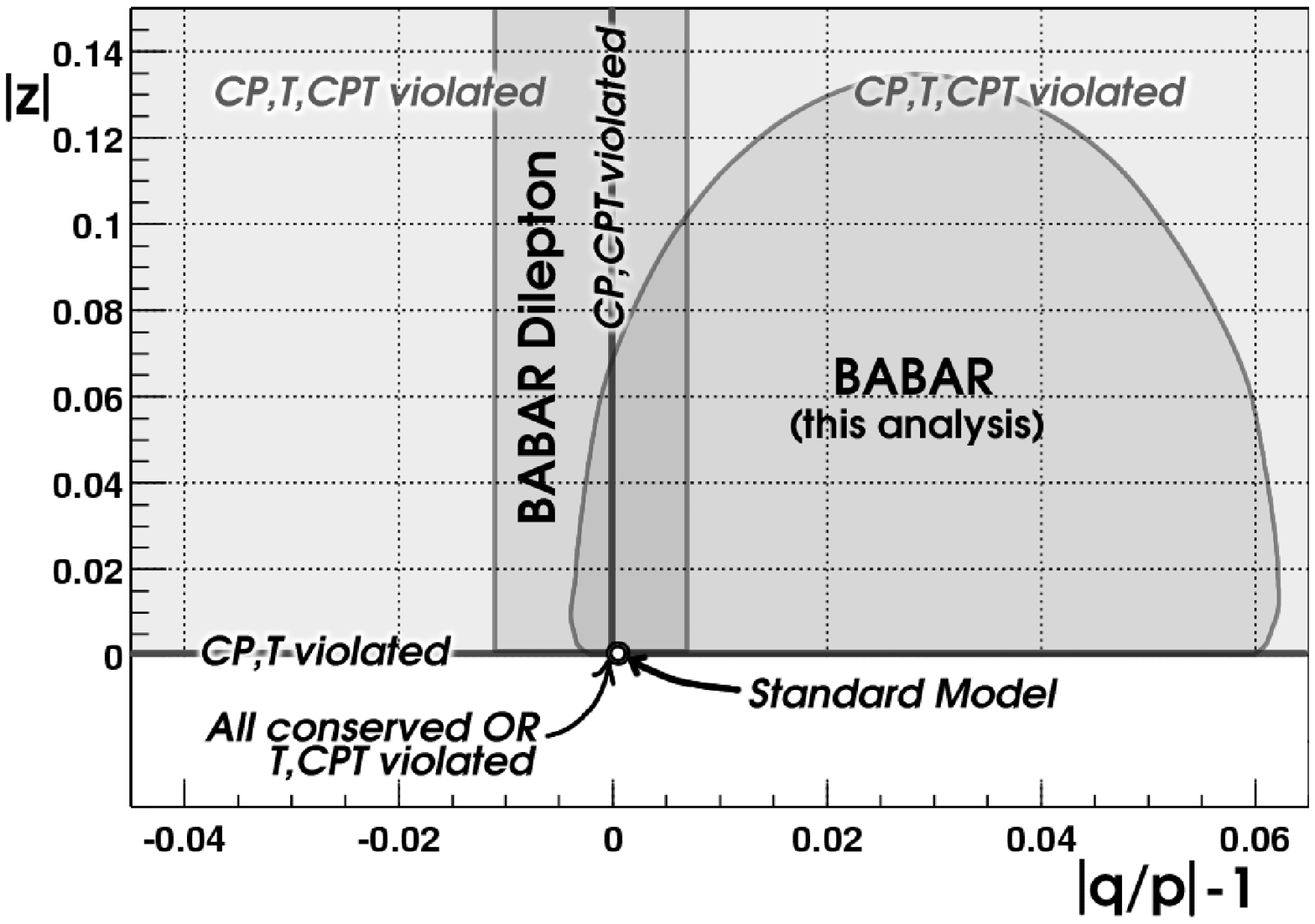,width=1.0\linewidth}
\caption{\fmvrc Favored regions at 68\% confidence level in the \mbox{$(\absqop-1,|\z|)$} plane
determined by this analysis and by the \babar\ measurement of 
the dilepton asymmetry~\cite{ref:babardileptonTviolation}. The axis labels reflect
the requirements that both \CP\ and \T\ be violated if $\absqop\ne 1$
and that both \CP\ and \CPT\ be violated if $|\z|\ne 0$. 
The Standard Model expectation for \absqop\ is obtained from
Refs.~\cite{ref:ciuchini,ref:absqopSM,ref:beneke}.}\label{fig:zqp}
\end{figure}

\noindent
The values in square brackets indicate the 90\% confidence-level intervals. 
When estimating the limits we also evaluate multiplicative contributions to the systematic error,
adding them in quadrature with the additive systematic uncertainties. 
{\fmvrc
Fig.~\ref{fig:zqp} shows the results in the $(\mbox{\absqop-1},|\z|)$ plane, compared to the 
\babar\ measurement of \absqop\ made with dilepton events, 
\mbox{$\absqop=0.998\pm0.006\pm0.007$}~\cite{ref:babardileptonTviolation}, and to 
the Standard Model expectations. 
The region shown for this analysis is obtained by simulating a large
number of experiments using the measured covariance matrix for the parameters
\reZ, \imZ, and \absqop, and is constrained to lie within the physical
region $|\z|\ge 0$. The three-dimensional distribution in $\reZ, \imZ$,
and $|q/p|$ is projected onto the two dimensions $|\z|^2$ and $|q/p|$.
The boundary is then chosen to exclude the maximal region.
For simplicity in the {\fmvcw figure, we} display $|\z|$ rather than $|\z|^2$.
The dilepton measurement constrains \absqop\ without assumptions on the value of
$|\z|$. The region in this case is obtained from the $\Delta \chi^2=1$ limits for this single variable.
}

Assuming \CPT\ invariance the results are
$$\begin{array}{l}
\sgndGoverG = \\
\resline{-0.009}{0.037}{0.018}{-0.085}{0.067}~,\\
\absqop = \\
\resline{\phantom{+}1.029}{0.013}{0.011}{\phantom{+}1.001}{1.057}~.
\end{array}$$
\begin{figure}[!t]
\centering
\begin{tabular} {c}  
\epsfig{file=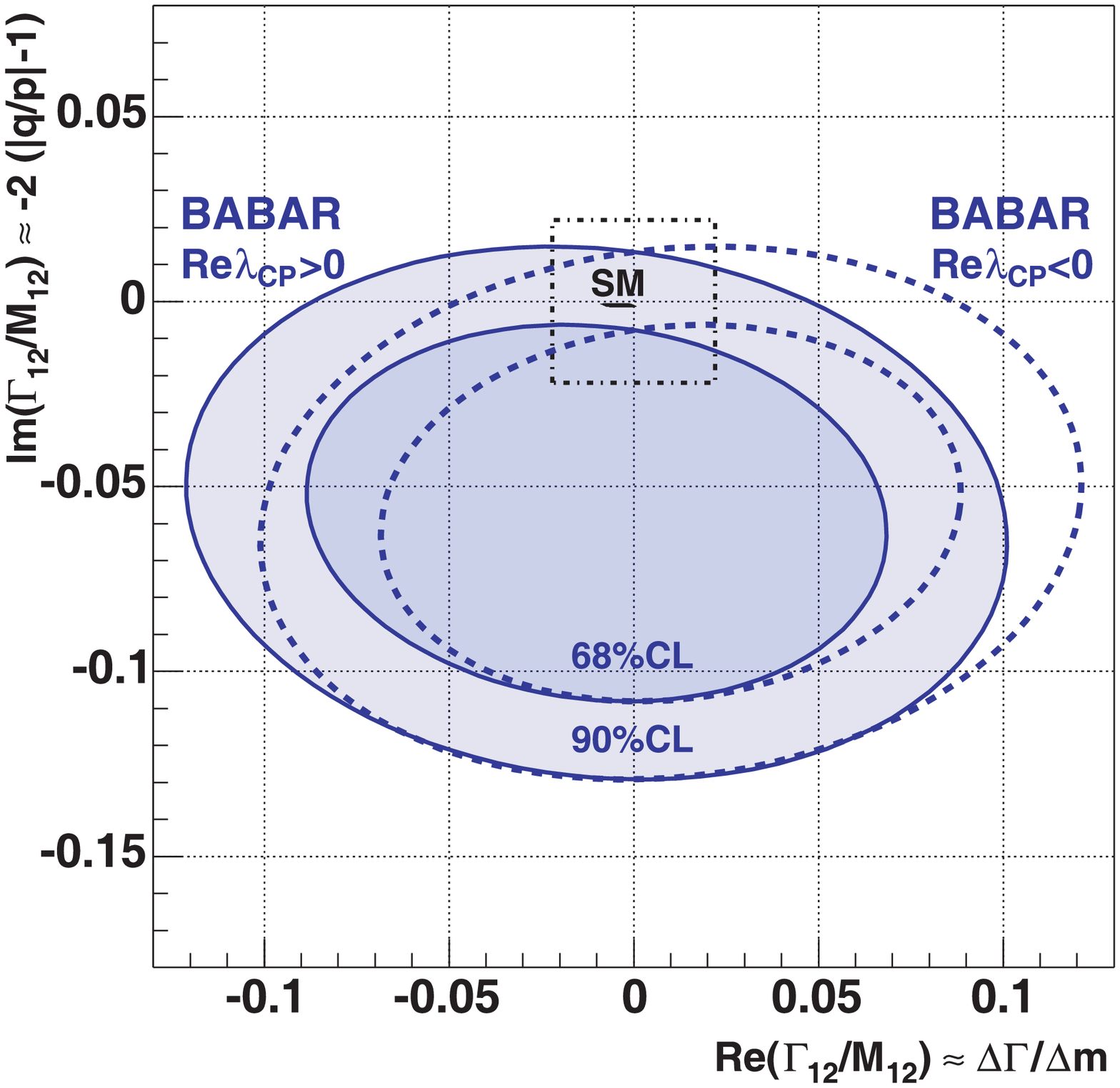,width=0.95\linewidth} \\
\epsfig{file=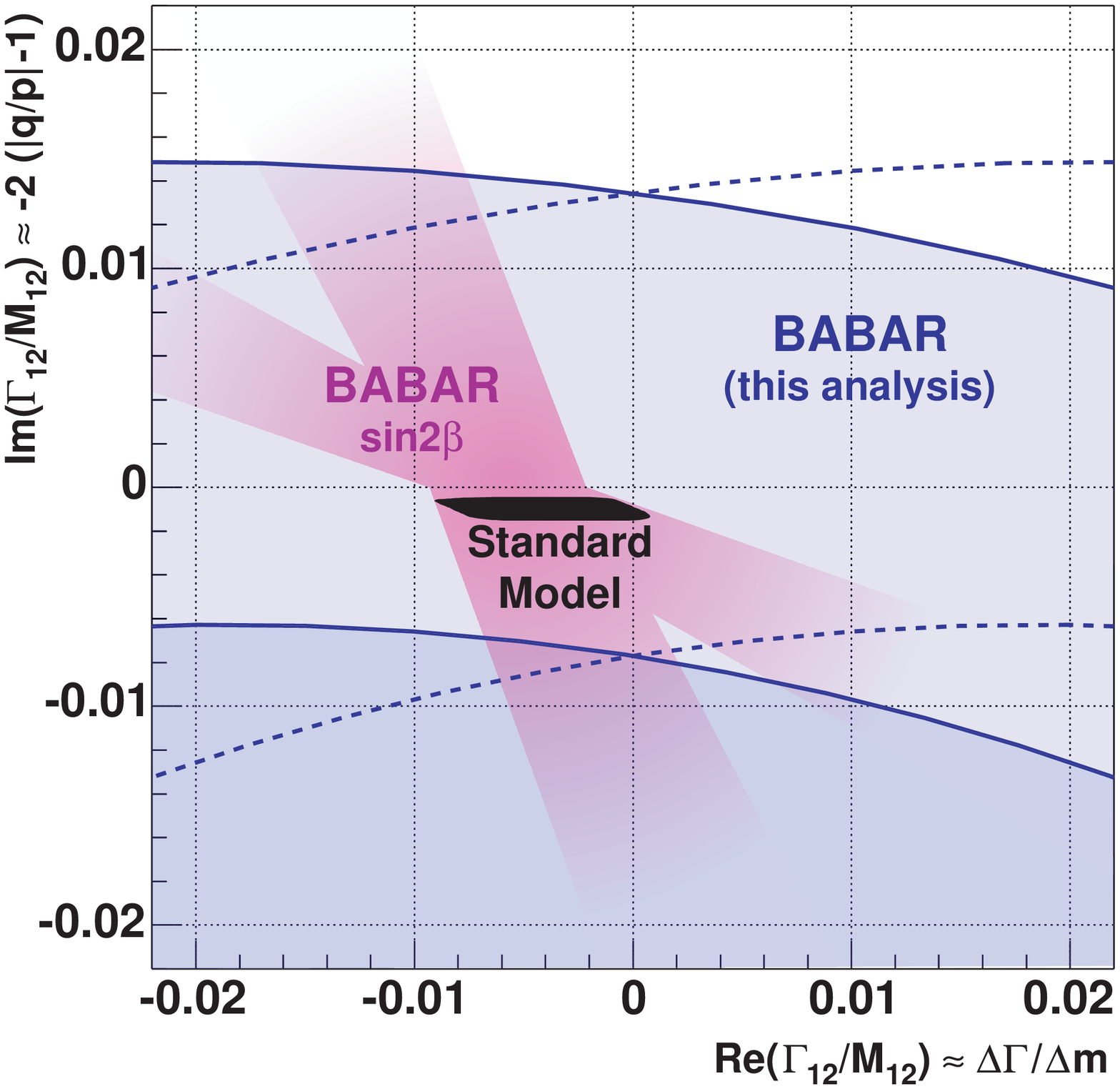,width=0.95\linewidth}
\end{tabular}   
\caption{\fmvrc Constraints at 68\% and 90\% confidence level on 
the complex ratio $\Gamma_{12}/M_{12}$ of the effective Hamiltonian 
off-diagonal matrix elements governing neutral-\B-meson oscillations
as determined from the \sgndGoverG\ and \absqop\ measurements 
of this analysis with \z\ fixed to zero, compared to predictions of Standard Model calculations when 
other experimental inputs are used.
The lower figure is an enlargement of the region around the origin.
\fmvrf{The bands in the lower figure are calculated using only the 
constraint obtained from the \babar\ $\sin2\beta$ measurement 
with \CP\ eigenstates like $\jpsi \KS$~\cite{ref:sin2b-babar}.
The fading out of the bands away from the origin indicates that these predictions 
are only valid for small $|\Gamma_{12}/M_{12}|$.}
}\label{fig:reGM_imGM}
\end{figure}

\noindent
{\fmvrc
These results can be used to set constraints on the complex ratio $\Gamma_{12}/M_{12}$ 
when \CPT\ invariance is assumed, 
as shown in Fig.~\ref{fig:reGM_imGM}.
Ellipses in the upper figure enclose the favored regions determined from the \sgndGoverG\ and \absqop\ measurements 
of this analysis with \z\ fixed to zero. Solid {\fmvcw contours} show the results assuming $\relambcpbare>0$
(as expected in the Standard Model based on other experimental constraints), while dashed
{\fmvcw contours} are for $\relambcpbare<0$. Inner (outer) contours represent 68\% (90\%) confidence-level regions 
for two degrees of freedom. 
The lower figure is an enlargement of the region around the origin of the 
complex $\Gamma_{12}/M_{12}$ plane.
The black region close to the origin of the complex plane
in the upper and lower figures shows the predictions of Standard Model calculations 
when all available experimental inputs are used to constrain the ratio of CKM matrix 
elements $(V_{cb}^{} V_{cd}^*)/(V_{tb}^{} V_{td}^*)$. 
The bands in the lower figure are calculated using only the constraint $\sin2\beta=0.741\pm0.075$ obtained from the
\babar\ measurement with \CP\ eigenstates like $\jpsi \KS$~\cite{ref:sin2b-babar}.
}

The decay-rate difference results can alternatively be expressed normalized to the 
mass difference \dM. {\fmvsix Using the world-average value of \dM~\cite{ref:pdg2002}, the result
allowing for \CPT\ violation (\z\ free) is}
$$\begin{array}{l}
\sgndGoverdM = \\
\resline{-0.011}{0.049}{0.024}{-0.112}{0.091}~,
\end{array}$$
and with \CPT\ invariance ($\z=0$)
$$\begin{array}{l}
\sgndGoverdM = \\
\resline{-0.012}{0.049}{0.024}{-0.113}{0.090}~.
\end{array}$$

The parameters \dM\ and \imlambcpflat\ are free in the fit, so that 
recent \B-factory \dM\ results~\cite{ref:dM-babar-had,ref:dM-babar-dstlnu,ref:dM-babar-dilep,ref:dM-belle}
and our $\sin2\beta$ analysis based on the same data sample~\cite{ref:sin2b-babar} provide a cross-check.
 The value of the \CP- and \T-violating parameter
\imlambcpflat\ increases by $+0.011$ when \CPT\ violation 
is allowed in the fit. \fmvrf{This change is equal to 15\% of the statistical uncertainty on
\imlambcpflat\ and is consistent with the correlations observed in the fit with \CPT\ violation.}

The results are consistent with Standard Model expectations and with
\CPT\ invariance.  To date, these are the 
{\fmvrc lowest}
limits on the
difference of decay widths of \Bz\ mesons and the strongest test of
\CPT\ invariance outside the neutral-kaon system~\cite{ref:cptkaons}.  
If we express the \CPT\ limits as ratios of the
\CPT-violating to the \CPT-conserving terms we have
\begin{eqnarray}
\frac{|\dMcpt|}{\M}<1.0\times 10^{-14}~,\nonumber\\
-0.156<\frac{\dGcpt}{\G}<0.042~\nonumber
\end{eqnarray}
at the 90\% confidence-level.
The limit on \CP\ and \T\ violation in mixing is
independent of and consistent with our previous measurement based on
the analysis of inclusive dilepton events~\cite{ref:babardileptonTviolation}. 
All the other results are also consistent with \fmvrf{previous
analyses~\cite{ref:dM-belle,ref:cleochid,ref:delphi,ref:otherCPTBtests,ref:pdg2002,ref:pdg2003,ref:belle_dilepton}.}  
All these measurements were obtained with more restrictive assumptions than those used here.
While the Standard Model predictions for \dG\ and
\absqop\ are still well below our current limits and no \CPT\ violation
is anticipated, higher precision measurements may still bring
surprises.

%% file: pubboard/acknowledgements.tex
We are grateful for the 
extraordinary contributions of our \pep2\ colleagues in
achieving the excellent luminosity and machine conditions
that have made this work possible.
The success of this project also relies critically on the 
expertise and dedication of the computing organizations that 
support \babar.
The collaborating institutions wish to thank 
SLAC for its support and the kind hospitality extended to them. 
This work is supported by the
US Department of Energy
and National Science Foundation, the
Natural Sciences and Engineering Research Council (Canada),
Institute of High Energy Physics (China), the
Commissariat \`a l'Energie Atomique and
Institut National de Physique Nucl\'eaire et de Physique des Particules
(France), the
Bundesministerium f\"ur Bildung und Forschung and
Deutsche Forschungsgemeinschaft
(Germany), the
Istituto Nazionale di Fisica Nucleare (Italy),
the Foundation for Fundamental Research on Matter (The Netherlands),
the Research Council of Norway, the
Ministry of Science and Technology of the Russian Federation, and the
Particle Physics and Astronomy Research Council (United Kingdom). 
Individuals have received support from 
the A. P. Sloan Foundation, 
the Research Corporation,
and the Alexander von Humboldt Foundation.

%% file: appendix.tex
The use of untagged data is essential for determining the asymmetries in
the tagging and reconstruction efficiencies.  To indicate how the various
samples enter we provide a simple example using only time-integrated quantities.
In practice we use a time-dependent analysis, which gives better precision because
it uses more information. 

Suppressing the indices for the tag category {\fmvsix index} $\alpha$ and the signal or background
component $j$, and writing the reconstruction efficiencies as
$\rho=\rho^j_{\flavv},\,{\overline \rho}=\rho^j_{\flavbarv}$ and the
tagging efficiencies as
$\tau=\tau^{\alpha,j}_{\tagv},\,{\overline \tau}=\tau^{\alpha,j}_{\tagbarv}$, Eq. (\ref{eq:numudef}) reads
\bea
 \nu & = & \frac{\rho-{\overline \rho}}{\rho+{\overline \rho}}~,\nn\\
 \mu & = & \frac{\tau-{\overline \tau}}{\tau+{\overline \tau}}~.
\label{eq:numudef2}
\eea
Using the numbers of signal events that are tagged and have a reconstructed \Bz ($X$), those
tagged and having a \Bzb ($Y$), those untagged with a reconstructed \Bz ($Z$), and finally
those untagged with a reconstructed \Bzb ($W$), we can determine the required asymmetries~\cite{ref:babar-stwob-prd}.  
To see this, note that if the total number of $\BzBzb$ pairs is $N$, and neglecting \dG, \mbox{$\absqop-1$}, 
and $\z$ corrections,
there are 
\bea
N_u & = &N(1+[1/(1+x^2)])/2
\eea
unmixed
events (i.e., $\BzBzb$) and  
\bea
N_m & = & N(1-[1/(1+x^2)])/2
\eea 
mixed events (i.e., $\Bz\Bz$ or $\Bzb\Bzb$), 
where $x=\dM/\G$. {\fmvrc Then we have}
\begin{eqnarray}
X&=&\rho \tau N_m/2 +\rho{\overline \tau}N_u/2\nonumber~,\\
Y&=&{\overline \rho}\hspace{1pt}{\overline \tau} N_m/2 +{\overline \rho}\tau N_u/2\nonumber~,\\
Z&=&\rho(1-\tau) N_m/2 +\rho(1-\overline \tau)N_u/2\nonumber~,\\
W&=&{\overline \rho}(1-\overline \tau) N_m/2 +{\overline \rho}(1-\tau) N_u/2~.
\end{eqnarray}
Setting $U=X+Z$ and $V=Y+W$, we find
\begin{equation}
\nu=\frac{U-V}{U+V}~~ ,~~~ \mu=(1+x^2)\frac{(Y/V)-(X/U)}{(Y/V)+(X/U)}~.
\end{equation}
Corrections to these equations have to be applied due to non-zero values of \dG, \mbox{$\absqop-1$} and $\z$.
The use of untagged events is {\fmvsix therefore} essential to the determination of $\nu$ and $\mu$.

%% file: paper.bbl
\begin{thebibliography}{99}

\bibitem{ref:dM-babar-had} \babar\ Collaboration, B.\ Aubert {\em et al.},
\jprl{88}, 221802 (2002).

\bibitem{ref:dM-babar-dstlnu} \babar\ Collaboration, B.\ Aubert {\em et al.}, 
\jprd{67}, 072002 (2003).

\bibitem{ref:dM-babar-dilep} \babar\ Collaboration, B.\ Aubert {\em et al.},
\jprl{88}, 221803 (2002).

\bibitem{ref:dM-belle} Belle Collaboration, N.C. Hastings {\em et al.}, 
\jprd{67}, 052004 (2003).

\bibitem{ref:sin2b-babar} \babar\ Collaboration, B. Aubert {\em et al.}, \jprl{89}, 201802 (2002).

\bibitem{ref:sin2b-belle} Belle Collaboration, K. Abe {\em et al.}, \jprd{66}, 071102(R) (2002).

\bibitem{ref:dighe} A.S.~Dighe, T.~Hurth, C.S.~Kim, and T.~Yoshikawa, \npb{624}, 377 (2002).

\bibitem{ref:ciuchini} M.~Ciuchini, E.~Franco, V.~Lubicz, F.~Mescia, and C.~Tarantino, JHEP {\bf 0308}, 031 (2003).

\bibitem{ref:cleochid} CLEO Collaboration, B.H.~Behrens {\em et al.}, \plb{490}, 36 (2000).  

\bibitem{ref:delphi} DELPHI Collaboration, J.~Abdallah {\em et al.}, \epjc{28}, 155 (2003).

\bibitem{ref:babardileptonTviolation}  \babar\ Collaboration, B.~Aubert {\em et al.}, \jprl{88}, 231801 (2002).


\bibitem{ref:absqopSM} S.~Laplace, Z.~Ligeti, Y.~Nir, and G.~Perez, \jprd{65}, 094040 (2002).

\bibitem{ref:beneke} M.~Beneke, G.~Buchalla, A.~Lenz, and U.~Nierste, \plb{576}, 173 (2003).

\bibitem{ref:sanda} M.~Kobayashi and A.I.~Sanda, \jprl{69}, 3139 (1992).

\bibitem{ref:ko} D.~Colladay and V.A.~Kosteleck\'y, \plb{344}, 259 (1995);
V.A.~Kosteleck\'y and R.~Van Kooten, \jprd{54}, 5585 (1996);
V.A.~Kosteleck\'y, \jprd{64}, 076001 (2001).

\bibitem{ref:bbCPT}{M.C.~Ba\~nuls and J.~Bernab\'eu,
\plb{464}, 117 (1999); \npb{590}, 19 (2000).}

\bibitem{ref:cpttheo} G.~L\"uders, Dansk. Math. Phys. Medd. {\bf 28}, 5 (1954); Ann. Phys. {\bf 2}, 1 (1957); 
W.~Pauli, \ncim{6}, 204 (1957); 
R.~Jost, Helv. Phys. Acta {\bf 30}, 409 (1957); 
F.J.~Dyson, \pr{110}, 579 (1958); 
R.F.~Streater and A.S.~Wightman, {\it PCT, Spin and Statistics, and All That}, Benjamin, New York, 1964.



\bibitem{ref:cptbreak} M.S.~Berger and V.A.~Kosteleck\'y, \jprd{65}, 091701 (2002);
V.A.~Kosteleck\'y and R.~Potting, \jprd{51}, 3923 (1995).

\bibitem{ref:cptkaons} \fmvrf{KTeV Collaboration, A.~Alavi-Harati {\em et al.}, \jprd{67}, 012005 (2003);}
CPLEAR Collaboration, A.~Apostolakis {\em et al.}, \plb{456}, 297 (1999);
\fmvrf{E773 Collaboration, B.~Schwingenheuer {\em et al.}, \jprl{74}, 4376 (1995);}
NA31 Collaboration, R. Carosi {\em et al.}, \plb{237}, 303 (1990).


\bibitem{ref:otherCPTBtests} OPAL Collaboration, R.~Ackerstaff {\em et al.}, Z. Phys. C {\bf 76}, 401 (1997).

\bibitem{ref:CPTprl} \babar\ Collaboration, B.~Aubert {\em et al.},
arXiv:hep-ex/0311037, \fmvrf{to appear in} \jprl{}

\bibitem{ref:owenandus} O.~Long, M.~Baak, R.~Cahn, and D.~Kirkby, \jprd{68}, 034010 (2003).

\bibitem{ref:grossman} Y.~Grossman, A.L.~Kagan, and Z.~Ligeti, \plb {\bf 538}, 327 (2002).

\bibitem{ref:cleojpsikp} CLEO Collaboration, G.~Bonvicini {\it et al.}, \jprl{84}, 5940 (2000).

\bibitem{ref:babarjpsikp} \babar\ Collaboration, B.\ Aubert {\em et al.}, 
\jprd{65}, 091101 (2002).

\bibitem{ref:theojpsikp} R.~Fleischer and T.~Mannel, \plb{506}, 311 (2001).  

\bibitem{ref:babar-nim} \babar\ Collaboration, B.~Aubert {\em et al.}, \nima{479}, 1 (2002).

\bibitem{ref:geant4} S.~Agostinelli {\em et al.}, \nima{506}, 250 (2003).


\bibitem{ref:babar-stwob-prd}{\babar\ Collaboration, B.~Aubert {\em et al.}, \jprd{66}, 032003 (2002).}

\bibitem{ref:argus} ARGUS Collaboration, H.~Albrecht {\em et al.}, Z. Phys. C {\bf 48}, 543 (1990).

\bibitem{ref:pdg2002} Particle Data Group, K.~Hagiwara {\em et al.}, \jprd{66}, 010001 (2002).

\bibitem{ref:pdg2003} \fmvrf{2003 partial update of Ref.~\cite{ref:pdg2002} for edition 2004 (http://pdg.lbl.gov).}

\bibitem{ref:belle_dilepton} Belle\ Collaboration, N.C.~Hastings {\em et al.}, \jprd{67}, 052004 (2003).










\end{thebibliography}
